%% file: main.tex
\documentclass[sigconf, nonacm]{acmart}

\AtBeginDocument{%
  \providecommand\BibTeX{{%
    \normalfont B\kern-0.5em{\scshape i\kern-0.25em b}\kern-0.8em\TeX}}}

\usepackage{algpseudocode}
\usepackage{algorithm}
\usepackage{amsmath}
\usepackage{xcolor}
\usepackage{subcaption}

\usepackage{multirow}
\usepackage{breqn}
\usepackage[scientific-notation=true]{siunitx}

\algnewcommand\algorithmicinput{\textbf{Input:}}
\algnewcommand\algorithmicoutput{\textbf{Output:}}
\algnewcommand\Input{\item[\algorithmicinput]}%
\algnewcommand\Output{\item[\algorithmicoutput]}%

\algnewcommand\algorithmicpinput{\textbf{\quad \ \ Input:}}
\algnewcommand\PInput{\item[\algorithmicpinput]}%

\MakeRobust{\Call}

\setcopyright{acmcopyright}
\copyrightyear{2018}
\acmYear{2018}
\acmDOI{10.1145/1122445.1122456}

\acmConference[Woodstock '18]{Woodstock '18: ACM Symposium on Neural
  Gaze Detection}{June 03--05, 2018}{Woodstock, NY}
\acmBooktitle{Woodstock '18: ACM Symposium on Neural Gaze Detection,
  June 03--05, 2018, Woodstock, NY}
\acmPrice{15.00}
\acmISBN{978-1-4503-XXXX-X/18/06}

\newcommand{\songadd}[1]{\textcolor{red}{#1}}
\definecolor{burntorange}{rgb}{0.8, 0.33, 0.0}

\begin{document}

\abovedisplayskip=1pt
\abovedisplayshortskip=1pt
\belowdisplayskip=1pt
\belowdisplayshortskip=1pt
\abovecaptionskip=3pt
\belowcaptionskip=3pt

\setlength{\textfloatsep}{5pt}

\title{OpenGraphGym-MG: Using Reinforcement Learning to Solve Large Graph Optimization Problems on MultiGPU Systems\\}

\author{Weijian Zheng}
\affiliation{%
  \institution{Indiana University-Purdue University} 
  \streetaddress{723 W Michigan St, SL 280}
  \city{Indianapolis}
  \country{USA}}
\email{zheng273@purdue.edu}

\author{Dali Wang}
\affiliation{%
  \institution{Oak Ridge National Laboratory}
  \city{Oak Ridge}
  \country{USA}
}
\email{wangd@ornl.gov}

\author{Fengguang Song}
\affiliation{%
  \institution{Indiana University-Purdue University} 
  \streetaddress{723 W Michigan St, SL 280}
  \city{Indianapolis}
  \country{USA}}
\email{fgsong@iupui.edu}


\begin{abstract}

Large scale graph optimization problems arise in many fields. 
This paper presents an extensible, high performance framework (named {\it OpenGraphGym-MG}) 
that uses deep reinforcement learning and graph embedding to 
solve large graph optimization problems with multiple GPUs. 
The paper uses a common RL algorithm (deep Q-learning) and a representative graph embedding (structure2vec) to demonstrate the extensibility of the framework and, most importantly, to illustrate the novel optimization techniques, 
such as spatial parallelism, graph-level and node-level batched processing, distributed sparse graph storage, efficient parallel RL training and inference algorithms, repeated gradient descent iterations, and adaptive multiple-node selections.
This study performs a comprehensive performance analysis on parallel efficiency and memory cost that proves the parallel RL training and inference algorithms are efficient and highly scalable on a number of GPUs. This study also conducts a range of large graph experiments, with both generated graphs (over 30 million edges) and real-world graphs, using a single compute node (with six GPUs) of the Summit supercomputer. Good scalability in both RL training and inference is achieved: as the number of GPUs increases from one to six, the time of a single step of RL training and a single step of RL inference on large graphs with more than 30 million edges, is reduced from 316.4s to 54.5s, and 23.8s to 3.4s, respectively. The research results on a single node lay out a solid foundation for the future work to address graph optimization problems with a large number of GPUs across multiple nodes in the Summit.
\end{abstract}




\keywords{Reinforcement learning, big graph optimization problems, distributed GPU computing, open AI software environment}

\maketitle

\input{sections/1_introduction}
\input{sections/2_related_work}

\input{sections/3_open_RL_design}

\input{sections/4_0_implementations}

\input{sections/5_analysis}

\input{sections/6_experiment}

\input{sections/7_conclusion}

\begin{acks}
This work is based upon research supported by the U.S. Department of Energy, Office of Science, Advanced Scientific Computing Research. This research used resources of the Oak Ridge Leadership Computing Facility at the Oak Ridge National Laboratory. 
\end{acks}

\bibliographystyle{ACM-Reference-Format}
\bibliography{paperlist}

\end{document}

%% file: sections/1_introduction.tex
\section{Introduction}
\label{sec:intro}

The study of graph optimization problems is one of the core areas in computer science with a wide variety of applications in social sciences, operations research, power systems, chemistry, and bioinformatics. Many of the graph problems are NP-hard and intractable. 
To solve the problem, in practice, researchers have designed three categories of algorithms: {\it exact}, {\it approximation}, and {\it heuristic} algorithms.
The {\it exact} algorithms find optimal solutions but only work for small-size graphs or graph problems with fixed parameters \cite{FPT}.
The {\it approximation} algorithms have polynomial time complexities and provide an upper bound for their approximation ratios, but their solutions are often significantly worse than the optimal solutions.
The {\it heuristic} algorithms use different heuristics to search for optimal solutions, and can usually find high-quality solutions with faster performance than the approximation algorithms.

The {\it heuristic} algorithms work well in practice, but these algorithms often require extensive expert and domain-specific knowledge.  
Recently, a few works have applied Reinforcement Learning (RL) to explore a large set of heuristics without expert/domain knowledge, and shown promising results. 
Examples of state‐of‐the‐art work include structure2vec \cite{song_le_paper},
GraphSage \cite{graphsage}, ECO-DQN \cite{barrett2020exploratory}, OrGym \cite{or-gym}, OpenGraphGym~\cite{opengraphgym}, and Ecole \cite{ecole}.
However, the existing RL work on graph optimizations either utilizes one GPU for RL, or does not scale well to address big graph optimization problems arising from real world applications (e.g., an Amazon graph can have 8.6 million nodes and 231.6 million edges \cite{10.1145/2766462.2767755}). 


In this paper, we design and implement a parallel MultiGPU-enabled Graph RL framework, named 
{\it OpenGraphGym-MG}, to provide an open AI environment dedicated to the area of applying RL to graph optimization problems.
The framework is able to
support parallel big graph RL training and inference on MultiGPU systems, and deliver high parallel efficiency and fast convergence rates with good-quality solutions.
This work particularly targets large-scale graphs that cannot fit in a single GPU's global memory, and designs both parallel RL training and parallel RL inference algorithms with minimized memory footprint
for handling big graphs. 
We also develop several optimization techniques to further improve the performance of the RL training and inference. 

We use the classic and fundamentally challenging Minimum Vertex Cover problem---which is one of Karp's 21 NP-complete problems~\cite{karp1972reducibility}---as an example to introduce our MultiGPU parallel graph RL algorithms, framework implementation, and parallel performances.
The policy model in our RL agent is based on the
widely used message-passing graph embedding~\cite{gilmer2017neural}. 
Specifically, we adopt a popular message-passing graph embedding model (i.e. {\it structure2vec} \cite{song_le_paper}) and a common RL algorithm (deep Q-network). However, the open design of our graph RL framework can be readily applied to incorporate other graph embedding models and different RL algorithms to solve various graph problems.

An analytical performance analysis shows that our parallel RL training algorithm and RL inference algorithm have a {\it parallel efficiency} that is close to $1.0$ when using parallel processes with multiple CPUs and GPUs.  
A memory cost analysis shows that our distributed data structures have a minimal memory overhead, 
and can minimize the amount of memory needed during big graph RL training.  Moreover, we perform experiments
on a compute node in the Summit supercomputer with two types of generated graph datasets and three real-world graphs. 
We also evaluate the learning speed, the quality of graph solutions, the advantages of using the new optimization techniques, and the execution time using multiple Nvidia V100 GPUs.
Our experimental results demonstrate that the OpenGraphGym-MG framework can provide good-quality solutions with an excellent computing performance, such as fast RL convergence rate, and nearly linear scalability in the RL training and inference over multiple GPUs.

In the remainder of the paper, next section presents the related work. Section \ref{sec:seq_alg} introduces the background of our open graph RL framework and challenges.
Section \ref{sec:par_alg} introduces the design and implementation of the OpenGraphGym-MG framework.
Section~\ref{sec:analysis} provides the analytical performance analysis and memory cost analysis for the parallel algorithms.
Finally, Section \ref{sec:exp} and Section \ref{sec:conclusion} present the experimental results and summarize the paper.

%% file: sections/2_related_work.tex
\section{Related work}
\label{sec:related_work}



There is a line of works using reinforcement learning for solving optimization problems, however, these works only support RL on a single GPU and do not aim to develop an extensible high performance graph RL framework using multiple GPUs. For example,
Song~et~al.~\cite{song_le_paper} use the structure2vec graph embedding and deep Q-network 
to solve the problems of Maximum Cut, Minimum Vertex Cover (MVC), and Travelling Salesman Problem (TSP).
Barrett et al.~\cite{barrett2020exploratory} develop the ECO-DQN method to allow {\it exploration at test time} to solve Maximum Cut.
Tang et al. \cite{pmlr-v119-tang20a} deploy
RL to adaptively select {\it cuts} to solve Integer Programming.
Other works include combining RL with pointer networks \cite{rl_graph_tsp}, and with graph convolutional networks (GCN)  \cite{rl_cornell,2019_gcn} to solve TSP, Minimum Spanning Tree, and Influence Maximization, respectively.
As RL on combinatorial optimizations (CO) has gained more attention, 
OpenGraphGym~\cite{opengraphgym}, OR-Gym~\cite{or-gym},  and Ecole \cite{ecole} have also been developed to study RL algorithms for solving different optimization problems.


It is also worthy to mention that many popular RL frameworks do not natively support graph data, and cannot be directly used to address graph optimization problems.
For instance, OpenAI Gym~\cite{openai_gym} provides a collection of environments for studying RL algorithms,
such as Atari, MuJoCo, Robotics, and some third-party environments. 
CuLE (CUDA Learning Environment) \cite{cule} 
is a CUDA port of the Atari learning environment, and supports Atari on multiple GPUs.
Horizon (or ReAgent) \cite{gauci2018horizon} 
is an end-to-end RL workflow framework that includes data preprocessing, normalization of features, performance reports, etc.
There are also several Distributed Multi-Agent RL frameworks, which use multiple agents to explore multiple environments in parallel. 
PARL \cite{parl} is a distributed RL framework for CPU-based clusters.
Ray RLlib \cite{liang2017ray} and Acme \cite{hoffman2020acme}
provide a modular RL framework to support distributed multi-agent learning.
Ray RLlib is recently deployed to work on multiple nodes of Cray XC systems~\cite{ray_cray}.

There are also a lot of works that use supervised learning and graph neural network (GNN) libraries \cite{dgl,fast_repre,graphnets,neugraph,aligraph,10.1145/3419111.3421281,zheng2020distdgl,jia2020improving,10.5555/3433701.3433794} to solve graph related applications.
Lately, a few researchers start to employ supervised GNN learning to solve combinatorial optimization problems.
Li et al.~\cite{2018_gcn} use deep learning and guided tree-search to train a GCN model to solve Maximal Independent Set.
Abe et al.~\cite{abe2019solving} use 
Monte Carlo Tree Search (MCTS) 
to train GNN models to solve MVC and Maximum Cut.
Wilder et al. design the CLUSTERNET system~\cite{wilder2019end} that uses GCN models and a version of K-means clustering to solve combinatorial optimization problems.
Unlike our method that uses RL to explore the intrinsic features of graphs and solve the graph problem together, these works use supervised learning approach and require a large number of problem instances with known optimal solutions for GNN training.

In summary, although there are works that applied RL to combinatorial optimization problems, to the best of our knowledge, our study is the first work that designs and implements an extensible, high performance RL framework for graph optimizations using multiple GPUs.

%% file: sections/3_open_RL_design.tex
\section{Deep Reinforcement Learning for Graph Optimization Problems}
\label{sec:seq_alg}

In this section, we introduce the general idea of designing an open RL framework for solving graph optimization problems, and report the challenges to realize a high-performance RL framework that can efficiently train/inference large-scale graphs on multiple GPUs.

\begin{figure} [h] 
    \centering
	\includegraphics[width=0.47\textwidth]{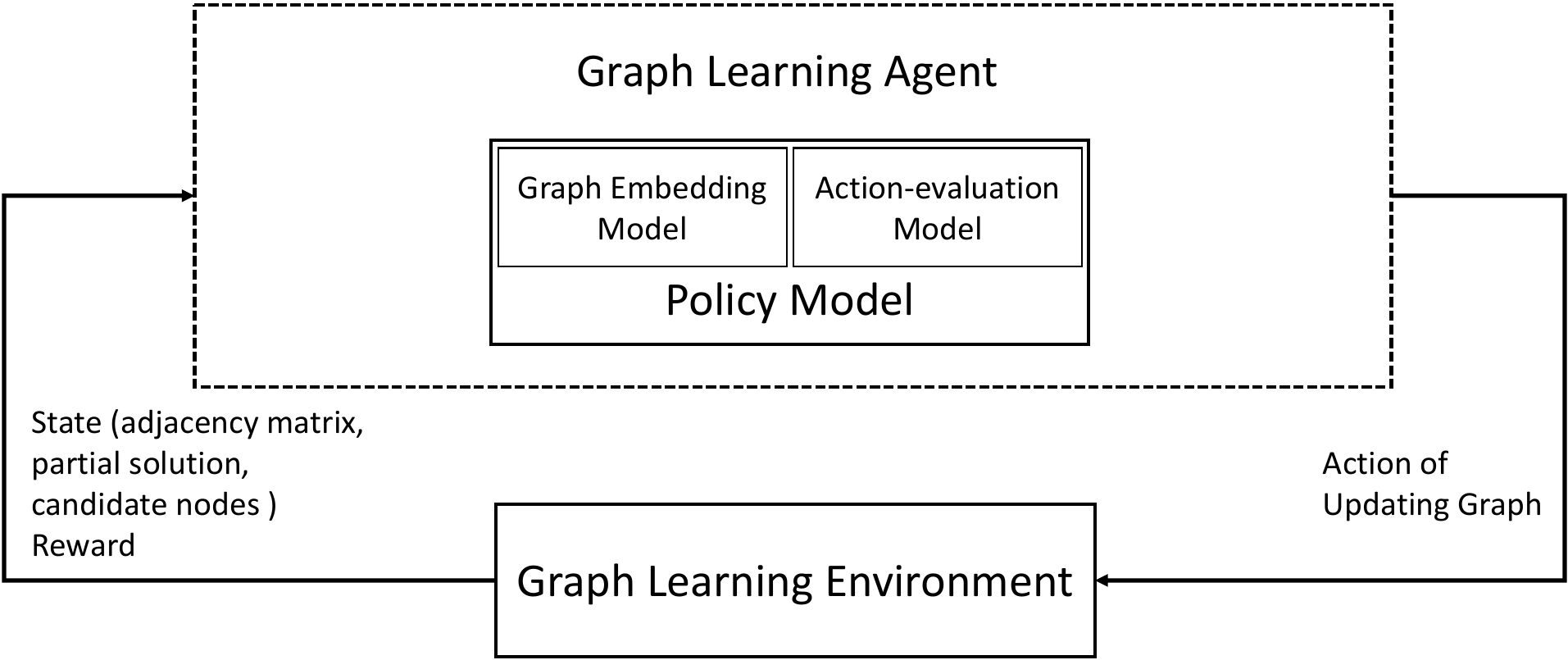}
	\par \centering
	\caption{\small An open modular graph RL framework, in which users can add new graph problem environments, try different RL algorithms, replace the graph embedding and action-evaluation models.
	}
	\label{fig:framework}
\end{figure}

Fig. \ref{fig:framework} shows the graph-problem-oriented open RL framework that allows users to apply different RL methods to solve various graph optimization problems.
The open graph RL framework consists of three major software modules: {\it Graph Learning Environment}, {\it Graph Learning Agent}, and {\it Policy Model}.

The Graph Learning Environment can modify a graph based on an ``action'' (i.e., graph manipulation operation) sent from the Graph Learning Agent. Then, it sends the new graph information: State, and Reward (due to taking the action), back to the Agent.
The Graph Learning Agent is employed to learn an optimal Policy Model that can find high-quality solutions to graph optimization problems by the trial-and-error learning.
The agent's policy model consists of a {\it graph embedding model} and an {\it action-evaluation model}. 
The {\it graph embedding model} is responsible for generating embedding vectors for a graph. 
The {\it action-evaluation model} takes the embedding vectors as input, and decides which node(s) should be selected and added to the partial solution. 
There exist a broad range of graph embedding models \cite{word2vec, structure2vec, deepwalk}, and we anticipate users will test and play with different graph embedding methods than the one used in this study. 

\textbf{A generic algorithm to use the open graph RL framework:} 
The interaction between the Graph Learning Agent and the Graph Learning Environment is essentially a {\it loop} structure.
We provide a simple, generic algorithm for using the graph RL framework,
as shown in Alg.~\ref{alg:rl_graph}.
In the algorithm, we first initialize a Graph Learning Agent.
The Graph Learning Agent can be configured with a user-defined policy model, which may consist of a user-defined embedding model and a user-defined action-evaluation model.
Each \textit{episode} of the RL training starts with the agent trying to work on a new graph.
As shown in Line 6 of Alg.~\ref{alg:rl_graph}, 
the RL framework randomly picks a new training graph $g$. 
Next, it creates a new graph learning environment $Env$ with the random graph $g$, as well as specifying the name of the graph optimization problem. 

\begin{algorithm}[tb]
	\caption{Pseudocode to use the Open Graph RL Framework.}
	\label{alg:rl_graph}
	\small{
		\begin{algorithmic}[1]
			
			\Input Graph\_Dataset: a list of training graphs
			\Statex \quad \ \ $EM$: graph embedding model (to be trained)
			\Statex \quad \ \ $Q$: action-evaluation model (to be trained)
			
			\State $B$: mini-batch size
			\State $R$: replay memory buffer size
			\State $/*$ Create an RL agent using the policy model EM and Q $*/$
			\State Agent $\xleftarrow{}$ \Call{Graph\_Learning\_Agent}{EM, Q, Replay\_Buffer\_Size = R} 

			\For{each $episode$ $e$}

		    \State Randomly pick a graph ${g}^{}$ from \textit{Graph\_Dataset} 
		    \State $/*$ Create a new graph problem environment with graph $g$ $*/$
		    \State Env $\xleftarrow{}$ \Call{Graph\_Learning\_Env}{graph\_problem\_name, g} 

		    \For{each $step$ $t$} 
		    
            \State $ v_{t}^{}=\left\{
            \begin{array}{@{}ll@{}}
            & \text{Select a node randomly, or} \\
            & \text{Agent.\Call{Act}{Env.\Call{Get\_Current\_State}{\text{}}}}
            \end{array}\right.$
            
            \State reward, done = Env.\Call{Step}{$v_t$}
    
		    \State $/*$ Push the tuple to replay buffer $*/$
		  	\State Agent.\Call{Remember}{Env.\Call{Get\_Previous\_State}{\text{}}, $v_t$, reward, \par 
	        \hskip\algorithmicindent \qquad \qquad \qquad \quad Env.\Call{Get\_Current\_State}{\text{}}}
		    
		    \State $/*$ Sample a batch of tuples from the replay buffer $*/$
		    \State tuples\_batch = Agent.\Call{Sample}{size = B}
		    
		    \State $/*$ Apply multiple iterations to train $EM$ and $Q$ $*/$
		    \State Agent.\Call{Train}{tuples\_batch}
		    
		    \If{done} break
		    \EndIf
			
			\EndFor
			\EndFor 
		
		\end{algorithmic}
	}
\end{algorithm}

For each \textit{step t} of an episode, the 
agent selects a node to be added to the partial solution either randomly (i.e., explore) or using the current policy model (i.e., exploit). 
After the environment applies the action $v_t$ (Line 11),
the agent receives a reward signal \textit{reward} and a
termination signal \textit{done} from the environment.
The agent then stores
a new {\it experience tuple} (previous state, action, reward, current state) to its \textit{replay buffer} (Line 13).
In order to train the policy model (i.e., $EM$ followed by $Q$), the agent samples a mini-batch of experience tuples from the replay buffer (Line 15). Then, the agent performs forward and backward propagations to train the policy model using the sampled mini-batch of tuples (Line~17).
After the current episode is finished, the RL training process starts a new episode by taking on a new training graph.

\textbf{The Challenges to Realize the Framework for Large Scale Graphs on Multiple GPUs:} 
Essentially, Alg.~\ref{alg:rl_graph} only provides a sequential version of the graph-oriented RL framework that can work on a single GPU.
However, in order to efficiently apply RL (Alg.~\ref{alg:rl_graph}) to large-scale graph optimization problems in parallel using multiple GPUs,
we must address the following challenges: 
\begin{itemize}
	\item{\textit{How to handle the big graph cases where a large graph requires more memory than a single GPU can provide?}}
	Due to the expensive memory consumption of big graph problems, we need more than one GPU to store big graphs efficiently.
	
	\item{\textit{How to design efficient parallel RL Training algorithms on multiple GPUs?}} 
	Parallel RL training algorithms 
	use a number of medium- to large-scale graphs to train an optimal policy model. The corresponding training process 	
	will be time consuming given a single GPU. Thus, we need to design a scalable parallel RL training algorithm on multiple GPUs.
	
	\item{\textit{How to design efficient  parallel RL Inference algorithms on multiple GPUs?}} 
	After we have trained an optimal policy model, users can use the trained model directly to solve large-scale graph optimization problems (i.e., RL inference). 
	Here, the unseen test graphs (one or multiple) may be at very large scales, and will require multiple GPUs to store and solve the graph(s) in parallel. To that end,
	we need to design a scalable parallel RL inference algorithm.
	
	\item{\textit{Optimization techniques to speed up both RL training and inference.}} To further reduce the time of the parallel RL training and inference processes, we also introduce several optimization techniques by modifying the parallel RL algorithms. 

\end{itemize}

	
    
	



%% file: sections/4_0_implementations.tex
\section{Design and Implementation of OpenGraphGym-MG}
\label{sec:par_alg}



In order to tackle these challenges,
we design and implement an open graph RL framework, which is named
{\it OpenGraphGym-MG} on MultiGPU systems. 
We first present our distributed data structures (Section \ref{subsec:mg_data_structure}).
Then, we present the parallel implementation of the graph-embedding and action-evaluation models (Section \ref{subsec:mg_model}). 
We also introduce the parallel RL inference and training algorithms, respectively (Sections \ref{subsec:mg_inference} and \ref{subsec:mg_training}). 
Finally, we present additional performance optimization techniques (Section \ref{subsec:mg_optimizations}).

Here, we use the classic Minimum Vertex Cover (MVC) problem as an example to show how the OpenGraphGym-MG framework is designed and developed. 
The MVC problem is defined as follows:
Given a graph $G = (V, E)$, find the smallest set of nodes $S \subseteq V$ such that every edge in $E$ is incident to at least one node in~$S$.

\input{sections/4_1_data_structures}

\input{sections/4_2_EM_model}

\input{sections/4_2_Qvalue_model}

\input{sections/4_3_parallel_infer}

\input{sections/4_4_parallel_train}

\input{sections/4_5_optimizations}

%% file: sections/4_1_data_structures.tex
\subsection{Distributed Data Structures}
\label{subsec:mg_data_structure}

OpenGraphGym-MG has three core data structures that are distributed across all GPUs:
1) adjacency matrices, 2) sets of candidate nodes, and 3) sets of partial solution. Each graph's state is represented by: one adjacency matrix $A$, one set of candidate nodes $C$, and one set of partial solution $S$. Given a number of $P$ GPUs, the graph's data $A$, $C$, and $S$ will be partitioned into $P$ partitions such that each GPU stores $1/P$-th of the data. We use {\it spatial parallelism} to process and compute large-scale graphs such that a single large-scale graph can be handled by $P$ GPUs.

\begin{figure} [b] 
    \centering
	\includegraphics[width=0.4\textwidth] 
	{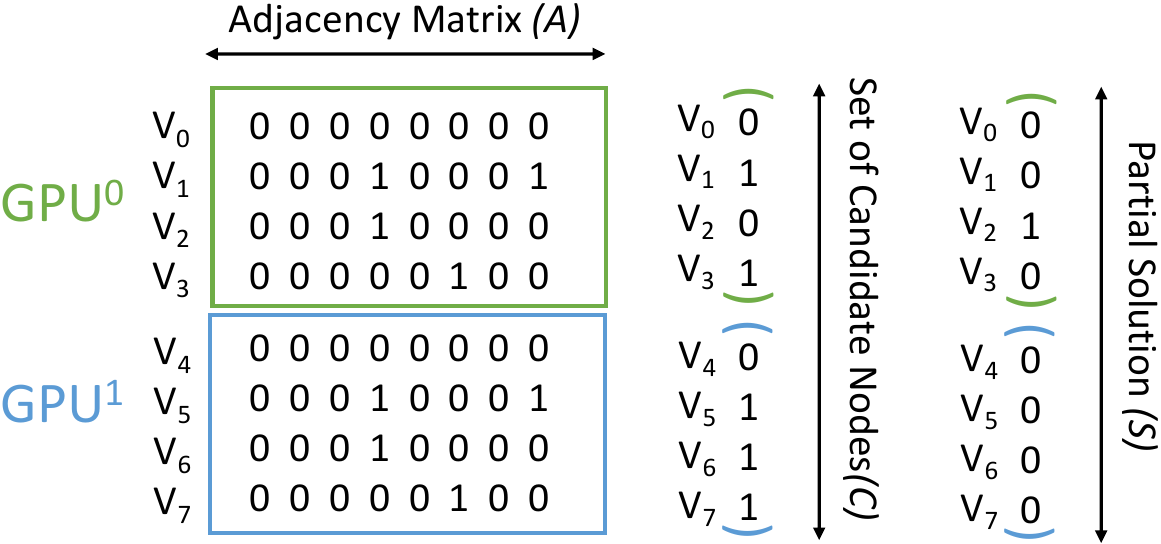}
	\par \centering
	\caption{\small{
		Three data structures (i.e., graph adjacency matrix $A$, a set of candidate nodes $C$, and current partial solution $S$) allocated on two GPUs (for an example of 8-node graph).
	}
	}
	\label{fig:single_graph_data_distribution}
\end{figure} 

As shown in Fig. \ref{fig:single_graph_data_distribution},
a graph's adjacency matrix $A$ is represented by an $N \times N$ matrix, where $N$ is the number of nodes in the graph ($N=8$ in the example).
Next, we partition the adjacency matrix along the row index, and distribute it to multiple GPUs. Each GPU has a $\frac{N}{P} \times N$ sub-adjacency-matrix, where $P$ is the number of GPUs ($P$ equals 2 in Fig. \ref{fig:single_graph_data_distribution}).  
The graph's set of candidate nodes $C$ is represented by a vector of $N$ elements. Each element $C_i$ is a binary number indicating whether node $V_i$ is still being considered as a candidate node or not. Only if a node is a candidate, the RL agent will evaluate its Q value to make the best decision. 
In Fig. \ref{fig:single_graph_data_distribution}, nodes $V_1$, $V_3$, $V_5$, $V_6$, and $V_7$ are the current candidate nodes. 
The partial solution set $S$ is also a vector of $N$ elements. $S_i=1$ indicates that node $V_i$ belongs to the current partial solution.

In addition, OpenGraphGym-MG particularly supports parallel training and parallel inference on a set of graphs in parallel.
To process a group of graphs in parallel, we stack the graphs together, and treat them as a 3D tensor. 
Although the graph data looks like image tensors used in image classifications, we choose not to use ``batch parallelism'' (aka ``data parallelism'') to distribute individual whole graphs to each GPU.
The main reason is that the size of a single graph may be too large to be stored or processed by one GPU, especially when users need to solve very large graph optimization problems. 
Hence, given a batch of $B$ graphs each with $N$ nodes, on $P$ GPUs, each GPU has a 
$B \times \frac{N}{P} \times N$ tensor for storing the graph batch's adjacency matrices,
a $B \times \frac{N}{P} \times 1$ tensor for the graph batch's candidate node sets,
and a $B \times \frac{N}{P} \times 1$ tensor for the partial solutions.

%% file: sections/4_2_EM_model.tex
\subsection{Implementation of the Graph Embedding and Action-Evaluation Models on Multiple GPUs}
\label{subsec:mg_model}


The RL agent's policy model is repeatedly evaluated by both RL training and inference processes.
This subsection introduces 
the parallel implementation of the RL agent's policy model. 
Our RL agent's policy model consists of two parts: 1) a graph embedding model, and 2) an action-evaluation model. The two models are connected into one ``combined'' model, which takes the current state of a graph,
and outputs scores for each node in the set of candidate nodes.
The methodology of message passing and information propagation has been widely used to solve graph related machine learning problems, so we use a recently-developed message passing model called {\it structure2vec} \cite{structure2vec} as an example to realize an RL agent. The other similar message passing based models \cite{message_gnn1, message_gnn2, gnn_survey} can also be added to OpenGraphGym-MG whenever needed.

Based upon the scores computed by the above policy model, the agent selects the node with the highest score, and adds it to the solution. The objective of connecting two models together is to tune and optimize both graph-embedding and action-evaluation models simultaneously to solve various graph optimization problems.

In the following content, we present the parallel algorithm to compute the graph embedding model (in Alg. \ref{alg:3d_parallel_em_model}), and the parallel algorithm to compute the action-evaluation model (in Alg. \ref{alg:3d_parallel_q_model}) on multiple GPUs, respectively.

Also, we let 
$B$ denote the size of a mini-batch of graphs,
$K$ denote the dimension of graph-embedding vectors,
$L$ denote the number of recurrent embedding layers,
$N$ denote the number of nodes of each graph,
$P$ denote the number of GPUs, and
$N^i = \frac{N}{P}$ denote the number of nodes allocated on the $i$-th GPU (GPU$^i$) in the paper.

\textbf{(1) Parallel Implementation of Graph Embedding Model}

The graph embedding model is a message-passing neural network model, where 
the embedding of node $v$ is updated recurrently with the information from its neighbor nodes $\in N(v)$. 
The message-passing embedding model can be expressed as follows \cite{structure2vec}:

\begin{equation}
\footnotesize{
\label{equa:emded}
embed_{v}^{L} = relu(\theta_{1}x_{v} + \theta_{4} 
\sum\limits_{u\in N(v)}^{} embed_{u}^{L-1} + \theta{}_{3} \sum\limits_{u\in N(v)}^{} relu(\theta{}_{2}(W(v, u)))),
}
\end{equation}

\begin{flushleft}
where $embed_v^L$ is the embedding of node $v$ at the $L_{th}$ embedding layer, 
$x_v$ is node $v$'s property,
$N(v)$ represents $v$'s neighbor nodes, and
$W(v, u)$ is the weight on the edge between $v$ and $u$.
Also, $\theta_{1}$, $\theta_{2}$, $\theta_{3}$, and $\theta_{4}$ are the model parameters used in the graph embedding model.
\end{flushleft}

Notice that Equation \ref{equa:emded} is only intended to express how to compute the embedding of a single node $v$.
Given many nodes, each node's embedding will be computed one at a time based on the mathematical equation. 
In practice, we need to reformulate the problem into matrix/tensor computations, 
and compute the embeddings for all the nodes and all the graphs in a batched manner. Here, we omit the mathematics details about how
to convert a single-node embedding computation to a batched multi-graph and multi-node embedding tensor computation.

We use Alg. \ref{alg:3d_parallel_em_model} to show the parallel algorithm to compute embeddings for a batch of $B$ graphs, each of which has $N$ nodes, based on Equation~\ref{equa:emded}.
Given $P$ GPUs, GPU$^i$ is allocated with a subset of the adjacency matrix tensor: $A^i$, and a subset of the partial solution tensor: $S^i$.
In Alg. \ref{alg:3d_parallel_em_model}, 
corresponding to Equation \ref{equa:emded},
we compute the term $\theta{}_{1}x_{v}$ (in Line 5), and term $\theta{}_{3} \sum\limits_{u\in N(v)}^{} relu(\theta{}_{2}(W(v, u)))$ (in Lines 7-8) for all the nodes located on GPU$^i$ for $B$ graphs.
The function {\it RESHAPE} is called to convert the parameters $\theta{}_{1} \in \mathbb{R}{}^{K \times 1} $, $\theta{}_{2} \in \mathbb{R}{}^{K \times 1}$, and $\theta{}_{3} \in \mathbb{R}{}^{K \times K}$ to $B \times K \times 1$, $B \times K \times N$, and $B \times K \times K$ dimensions to match different operands' dimensions to support batched tensor computations.
From Line 9 to Line 15, 
we compute the embeddings for GPU$^i$'s local resident nodes for $L$ iterations, which are implemented as $L$ layers in our graph embedding neural network.
In each iteration $l$, we first compute the sum of the embeddings of each node's neighbors (Lines 11-12). 
Next, we multiply the reshaped $\theta{}_{4}$ with 
the neighbor embedding sum
$nbr\_embed^i$. 
Then, we add $embed_1^i$,  
$embed_2^i$, and $embed_3^i$ to obtain the embeddings for all the nodes allocated on GPU$^i$.
Note that the embedding output on GPU$^i$ is a $B \times K \times N^i$ sparse tensor. 

\begin{algorithm}[b]
	\caption{Embedding Computation of \Call{EM}{$A^i$, $S^i$} on $\textbf{GPU}^{i}$}
	\label{alg:3d_parallel_em_model}
	\small{
		\begin{algorithmic}[1]
			
			\Input $A^{i} \in \mathbb{R}^{B \times N^i \times N}$: B adjacency matrices (in the sparse COO format)
			\Statex \quad \ \ $S^i \in \mathbb{R}^{B \times N^i \times 1}$: a stack of B partial MVC solutions for $B$ graphs  
            \State $/*$ Parameters of the model: $*/$
			\State $\theta{}_{1}, \theta{}_{2} \in \mathbb{R}{}^{K \times 1}$; $\theta{}_{3}, \theta{}_{4} \in \mathbb{R}{}^{K \times K}$
			
			\State $embed^i \in \mathbb{R}{}^{B \times K \times N^i}$: a stack of embeddings of nodes of the graphs, 
			\Statex \qquad \qquad \qquad \qquad \quad initialized to zero.
			
			\State $/*$ Transpose $S^i$ such that ${(S^i)}^T \in \mathbb{R}{}^{B \times 1 \times N^i}$ $*/$
			\State $embed_1^i$ = $\Call{MatMul}{\Call{reshape}{\theta{}_{1}, [B, K, 1]}, {(S^i)}^{T}}$    

    		\State $/*$ Transpose $A^i$ such that ${(A^i)}^T \in \mathbb{R}{}^{B \times N \times N^i}$ $*/$
    		\State $w^i$ = $\Call{ReLU}{\Call{SpMatMul}{\Call{reshape}{\theta{}_{2}, [B, K, N]}, {(A^{i})}^{T}}}$ 
			\State $embed_2^i$ = $\Call{MatMul}{\Call{reshape}{\theta{}_{3}, [B, K, K]}, w^i}$

			\For{\texttt{layer $l = 1$ to $L$}}
			
			\State $/*$ $nbr\_embed^i$ is the sum of  embeddings
			\Statex \qquad \ \  of each node's partial subset of neighbors on GPU$^i$ $*/$
			\State $nbr\_embed^i$ =  $\Call{SpMatMul}{embed^i, A^{i}}$ 
			
			\State $nbr\_embed^{}$ = $\Call{MPI-All-Reduce}{nbr\_embed^i, op = sum}$

			\State $embed_3^i$ = $\Call{MatMul}{\Call{reshape}{\theta{}_{4}, [B, K, K]}, nbr\_embed[i]}$  
 

			\State $embed^i$ = \Call{ReLU}{$embed_1^i$ + $embed_2^i$ + $embed_3^i$}
			\EndFor 
			
			\State $/*$ Return embedding for the local subset of nodes of the B graphs $*/$
			\State return $embed^i$ \Comment{$embed^i \in \mathbb{R}{}^{B \times K \times N^i}$}
            
		\end{algorithmic}
	}
\end{algorithm}



%% file: sections/4_2_Qvalue_model.tex
\textbf{(2) Parallel Implementation of Action-evaluation Model} ~

An action-evaluation model takes the local embeddings $embed^i$ and the local candidate node sets $C^i$ as input, and computes the scores for the candidate nodes $\in C^i$. This model is implemented as a neural network that has $\theta{}_{5}$, $\theta{}_{6}$, $\theta{}_{7}$ as model parameters.
The formula to compute the score for a candidate node $v$ is as follows: 

\begin{equation}
\label{equa:action_eva}
\footnotesize{
score_v=\theta^{T}_{7} relu[\theta{}_{5} \sum\limits_{u\in V}^{} embed{}_{u} ~||~ \theta{}_{6}embed^{}_{v}]
}
\end{equation}

\begin{algorithm}[t]
	\caption{Action-evaluation Computation of \Call{Q}{$embed^i$, $C^i$} on $\textbf{GPU}^{i}$}
	\label{alg:3d_parallel_q_model}
	\small{
		\begin{algorithmic}[1]
		    
			\Input $embed^{i} \in \mathbb{R}^{B \times K \times N^i}$: computed embedding of local subset of nodes 
			\Statex \quad \ \ $C_{}^{i} \in \mathbb{R}^{B \times N^i \times 1}$: a stack of candidate nodes for $B$ graphs

            \State $/*$ Parameters of the model: $*/$
			\State $\theta{}_{5}, \theta{}_{6} \in \mathbb{R}{}^{K \times K}$; $\theta{}_{7} \in \mathbb{R}{}^{2K \times 1} $
			
			\State $/*$ $sum\_embed^i$ is the sum of the local subset of nodes' embeddings $*/$
			\State $sum\_embed^i$ =  $\Call{Sum}{embed^{i}, axis=2}$ 
			
			\State $sum\_embed^{all}$ = $\Call{MPI-All-Reduce}{sum\_embed^i, op = sum}$ 

			\State $w_1^i = \Call{MatMul}{\Call{reshape}{\theta{}_{5}, [B, K, K]}, sum\_embed^{all}}$

            \State $/*$ convert each $C^i$ into a sparse diagonal matrix $*/$
            
			\State $candidate\_embed^i$ = $\Call{SpMatMul}{embed^{i}, \Call{sparse\_diag}{C^i}}$
            
			\State $w_2^i = \Call{MatMul}{\Call{reshape}{\theta{}_{6}, [B, K, K]}, candidate\_embed^i}$
			

			\State $w_3^i = \Call{ReLU}{\Call{Concat}{\Call{reshape}{w_1^i, [B, K, N^i]}, w_2^i}}$
			
			\State $scores^i = \Call{MatMul}{\Call{reshape}{{\theta{}_{7}}^{T}, [B, 1, 2K]}, w_3^i}$

			\State return $scores^i$ \Comment{$scores^i \in \mathbb{R}{}^{B \times 1 \times N^i}$}
			
		\end{algorithmic}
	}
\end{algorithm}

We use $[a ~||~ b]$ to denote a concatenation operation to concatenate two vectors to a longer vector.
The above equation is intended to compute for a single node's score instead of many nodes.
In practice, we need to compute scores for many candidate nodes from many graphs.
In Alg.~\ref{alg:3d_parallel_q_model}, we describe how each GPU$^i$ computes the scores for $B$ sets of local candidate nodes from $B$ graphs in parallel. 
First, each GPU computes the sum of all nodes' embeddings in every graph (Lines 4-5), which corresponds to the term $\sum\limits_{u\in V}^{} embed{}_{u}$ in Equation~\ref{equa:action_eva}. 
Then each GPU computes the product of $\theta{}_{5} \sum\limits_{u\in V}^{} embed{}_{u}$ (Line 6). 
In Line 8, we utilize a sparse diagonal matrix (constructed from the candidate node sets $C^i$), to extract the embeddings of those nodes that just belong to $C^i$ via a matrix computation.
Then, Line 9 computes $\theta{}_{6}embed^{}_{v}$
for all the candidate nodes. 
Line 10 performs a concatenation,
and applies \textit{ReLU} to it. 
Finally, each GPU obtains the scores for all the local candidate nodes resident on GPU$^i$ by multiplying $\theta{}_{7}$ with 
the \textit{ReLU} output (Line 11).

{\it Remark}: Regarding the communication time,
Alg. \ref{alg:3d_parallel_em_model} runs on $P$ GPUs and invokes a number $L$ of {\it MPI\_All\_reduce} communications, for which each GPU sends and receives a $B \times K \times N $ matrix. 
Alg.~\ref{alg:3d_parallel_q_model} has one {\it MPI\_All\_reduce} communication, for which each GPU sends and receives a $B\times K $ matrix. 
In Section~\ref{sec:analysis},
we will provide a detailed performance analysis for the two algorithms.

%% file: sections/4_3_parallel_infer.tex
\subsection{Parallel RL Inference Algorithm}
\label{subsec:mg_inference}


Based on the above algorithms that can evaluate the RL agent's policy model using $P$ GPUs, we are ready to introduce the parallel RL inference algorithm.
The RL inference algorithm utilizes the user-trained RL agent to seek a solution to an unseen test graph by following the agent's optimized policy model. 
Note the parallel RL {\it training} algorithm will be introduced in the next subsection. 

\begin{algorithm}[t]
	\caption{Parallel RL Inference on $\textbf{GPU}^{i}$}
	\label{alg:parallel_rl_testing}
	\small{
		\begin{algorithmic}[1]
		    \Input $EM$: graph embedding model (user-pretrained)
			\Statex \quad \ \ $Q$: action-evaluation model (user-pretrained)
			
			\Statex \quad \ \ $V_{}^{i}$: local subsets of nodes of the testing graphs on GPU$^i$

			\Statex \quad \ \ $A_{}^{i}$: local subsets of the adjacency matrices of the test graphs 
			
		    \State $S_{}^i$ = $\emptyset$: initial partial MVC solutions on GPU$^i$ 
		    \State $C_{}^{i}$: initial subset of candidate nodes in the test graphs from $V^i$
			
			\For{\texttt{step $t = 0$ to $|V|-1$}} \Comment{$|V|$: number of nodes in each graph}
		    
		    \State $embed^i = \Call{EM}{A_{}^{i}, S^i}$      
		    \Comment{$A^i \in \mathbb{R}{}^{B \times N^i \times N}$, sparse matrix}
		    \State $scores^i = \Call{Q}{embed^i, C_{}^{i}}$  
		    \State $scores^{all} = \Call{MPI-All-gather}{scores^i}$ 
		    \Comment{$scores^i \in \mathbb{R}{}^{B \times 1 \times N^i}$}

		    \State $v_{t}^{}= \text{$argmax_{v\in{V}}^{}   
            scores^{all}$}$
            \Comment{$scores^{all} \in \mathbb{R}{}^{B \times 1 \times N}$}
    
		    \State $S^i$ += $v_{t}^{}$; 
		    \State $C_{}^{i}$ -= $v_{t}^{}$; 
		    
		    \State $A_{}^{i} \xleftarrow{} \text{ Update local } A_{}^{i} \text{ by removing edges connected to } v_t$  
		    
		    \State if a graph solution is complete then break

			\EndFor

		\end{algorithmic}
	}
\end{algorithm}

Alg. \ref{alg:parallel_rl_testing} shows the parallel RL inference algorithm executed on each GPU$^i$. 
The algorithm takes as input one or more test graphs, and uses the pretrained embedding model $EM$ and action-evaluation model $Q$ to search for an optimal solution for each graph. Each GPU stores an instance of the pretrained $EM$ and $Q$ models.
When the algorithm starts, it first initializes the local partial-solution sets $S^i \in \mathbb{R}{}^{B \times N^i \times 1}$, where $B$ is the number of test graphs.
Then it initializes the local sets of candidate nodes $C^i$ $\in B\times N^i \times 1$ based on the local graph nodes $V^i$. 


The RL inference algorithm takes up to $|V|$ steps to find an optimal solution. 
In each step $t$, GPU$^i$ uses the pretrained $EM$ and $Q$ models to compute scores for its local resident candidate nodes (Lines 4-5) by using the previous Alg. \ref{alg:3d_parallel_em_model} and Alg. \ref{alg:3d_parallel_q_model}.
After gathering all scores, the candidate node with the highest (or top) score $v_t$ for each graph will be selected to be added to the partial solutions, meanwhile $S^i$ and $C^i$ are updated accordingly on GPU$^i$ (Lines 8-9). The local adjacency matrix $A^i$ is also updated based on the new action of selecting $v_t$. 
The RL inference process will finish when a complete graph solution is found.

\begin{figure} [b] 
    \centering
	\includegraphics[width=0.36\textwidth]{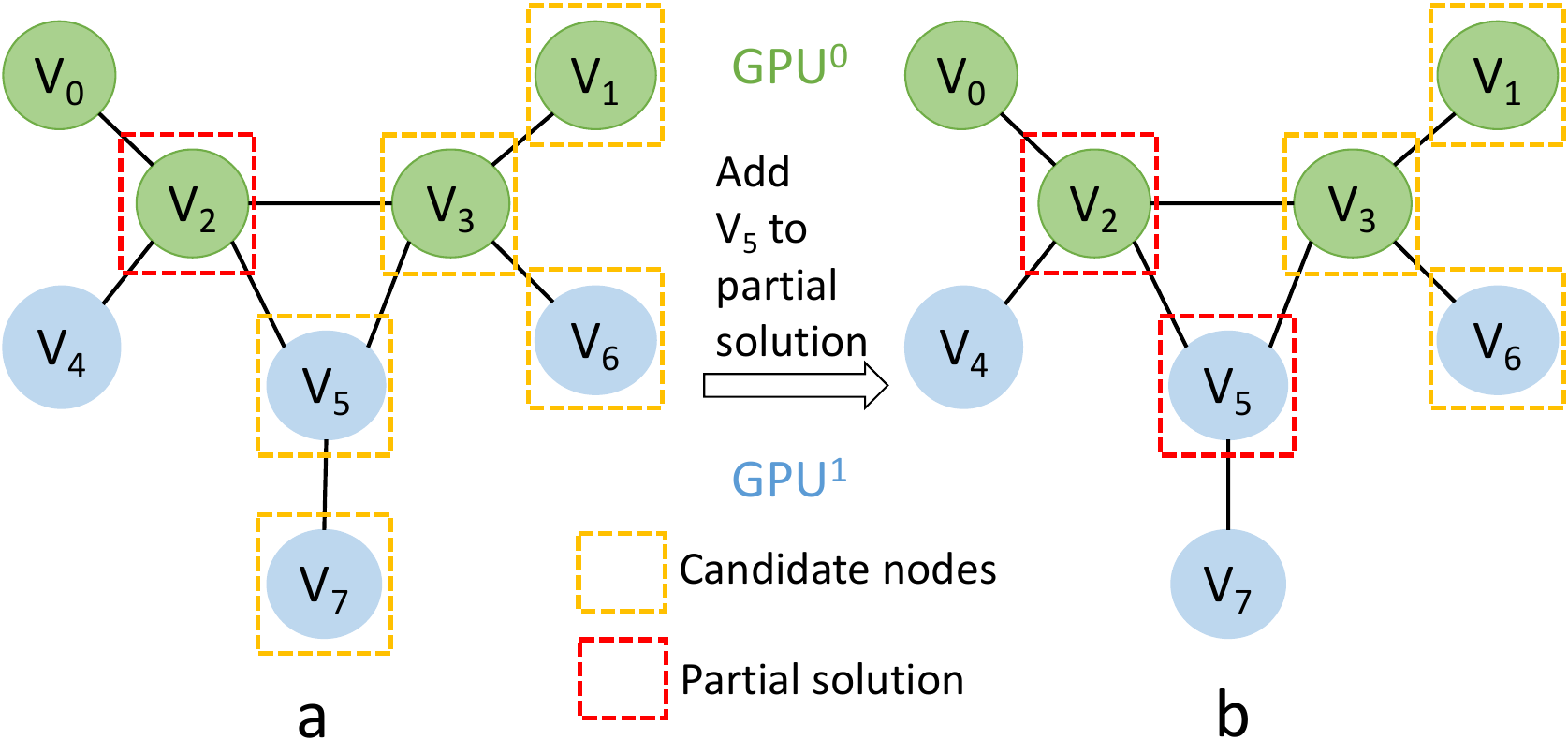}
	\par \centering
	\caption{\small{
	An example of selecting a new node.
	In (a), $V_2$ is in the partial solution. 
	In (b), $V_5$ is selected and added to the partial solution, $V_5$ and $V_7$ are afterwards removed from the candidate nodes.}
	}
	\label{fig:graph_example}
\end{figure}

\begin{figure} [t] 
    \centering
	\includegraphics[width=0.5\textwidth]{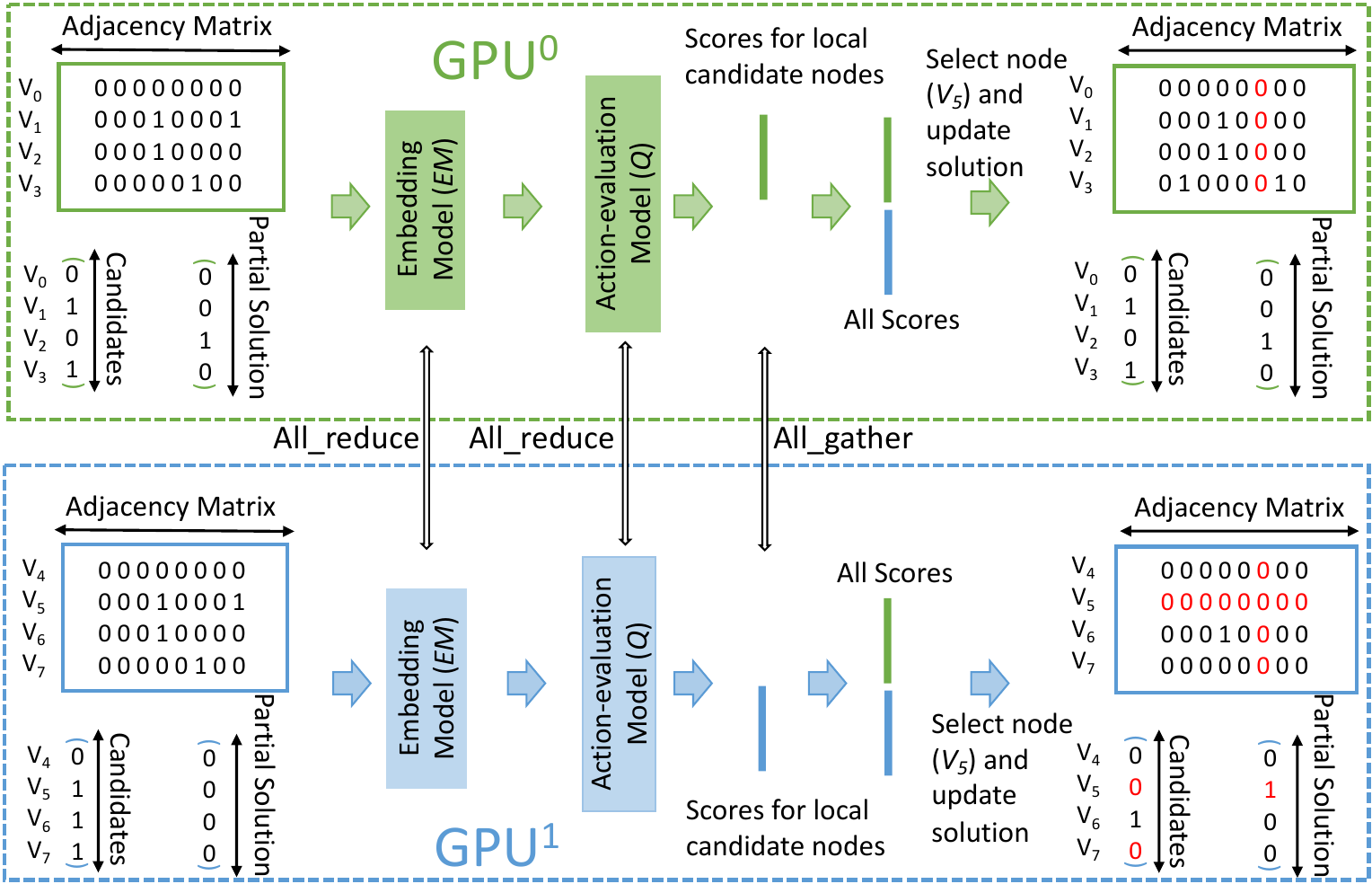}
	\par \centering
	\caption{\small{
	An illustration of one step of the OpenGraphGym-MG parallel inference algorithm using an 8-nodes graph. 
	Before this step, there are five candidate nodes as shown in Fig. \ref{fig:graph_example}a. Node $V_2$ is in the partial solution. Each GPU has a copy of the user-pretrained policy model, which includes the graph embedding model ($EM$) and the action-evaluation model ($Q$).
	At the beginning of the step, each GPU takes the distributed adjacency matrix, the local set of candidate nodes, and the local partial solution as the inputs. 
	Then, each GPU computes the scores for the corresponding nodes.
	Next, the scores from all GPUs are gathered. 
	Finally, the node with the highest score (node $V_5$) is added to the partial solution. Corresponding rows and columns in the adjacency matrix, the set of candidate nodes, and the partial solution are also updated (marked by red color).   
	}
	}
	\label{fig:par_inference}
\end{figure}

We will use a simple example to show how 
Alg. \ref{alg:parallel_rl_testing} is executed to solve the Minimum Vertex Cover (MVC) problem for a graph with 8 nodes using 2 GPUs.
Fig. \ref{fig:graph_example} shows the status of the example graph's partial solution set (red boxes)
and candidate node set (orange boxes) on GPU$^0$ (green) and GPU$^1$ (blue), before and after selecting node $V_5$ as a part of the solution.
Corresponding to deciding $V_5$ as a new partial solution node, we use Fig.~\ref{fig:par_inference} to illustrate how 
Alg. \ref{alg:parallel_rl_testing} works internally to make that decision, and how it updates its core data structures.


In Fig.~\ref{fig:par_inference}, 
GPU$^0$ is allocated with a half of the input graph's adjacency matrix, candidate node set, and partial solution (i.e., $V_0-V_3$).
GPU$^1$ is allocated with the same types of data but for nodes $V_4-V_7$.
Given the local data structures stored on each GPU and the pretrained {\it EM} and {\it Q} models,
GPU$^0$  is able to compute the scores for its own candidate nodes: $V_1$ and $V_3$, meanwhile
GPU$^1$ is able to compute the scores for its local candidates: $V_5, V_6, V_7$.
After a {\it gather} communication, each GPU obtains
the scores for all five candidates.
Next, the node with the highest score (e.g., $V_5$) is selected and added to the partial solution. 
After $V_5$ is moved from the candidate set to the partial solution, each GPU updates its local graph data structures.
Shown at the right end of Fig. \ref{fig:par_inference}, the two GPUs set the particular $V_5$-th row, and the sixth column (corresponding to $V_5$) in their subsets of adjacency matrix to all zero's. 
The local candidate node set and partial solution set on each GPU are also updated. 

%% file: sections/4_4_parallel_train.tex
\subsection{Parallel RL Training Algorithm}~
\label{subsec:mg_training}

We use Alg. \ref{alg:parallel_rl_training} to show the parallel RL training algorithm, with which an agent can learn an optimal policy model.
Given $P$ GPUs, a number of $P$ processes will be launched. Each process Proc$^i$ will utilize both a CPU and a GPU.
All operations that occur on GPUs are labeled with ``on GPU$^i$'' in Alg. \ref{alg:parallel_rl_training}, 
The environment-related computation is executed on CPUs, and the agent's policy evaluation and policy training are executed on GPUs.
Each Proc$^i$ owns a copy of the agent's policy model (i.e., the $EM$ and $Q$ models).
All processes will work together to solve one graph problem instance at a time.

\begin{algorithm}[t]
	\caption{Parallel RL Training on $\textbf{Proc}^{i}$ (with CPU and GPU$^i$)}
	\label{alg:parallel_rl_training}
	\small{
		\begin{algorithmic}[1]
			\Input Graph\_Dataset: a list of training graphs
			
			\Statex \quad \ \ SEED: a random seed used by all processes
			\Statex \quad \ \ $EM$: graph embedding model (to be trained)
			\Statex \quad \ \ $Q$: action-evaluation model (to be trained)
			
			\State $B$: mini-batch size for experience tuples
			\State Replay\_Buffer = $\emptyset$: experience replay buffer
			
			\For{each $episode$ $e$}
		    \State Randomly pick a graph ${g}^{}$ from \textit{Graph\_Dataset} 
		    \State $Env \xleftarrow{} \text{ Create a new environment using graph } {g}^{} $
			\State $A_{}^{i}$: local partition of adjacency matrix of graph $g$ on GPU$^i$
			
		    \State $C_{}^{i}$: local candidate nodes of graph $g$ on GPU$^i$
		    \State $S^i$ = $\emptyset$: current local partial solution of graph $g$ on GPU$^i$

		    \For{each $step$ $t$} 
		    
		    \State
            $ v_{t}^{}=\left\{
            \begin{array}{@{}ll@{}}
            & \text{Randomly explore (select a node using same seed), or} \\
            & \text{Policy Model exploit (i.e., Q(EM($A^i$, $S^i$), $C^i$))} \textrm{ on GPU}^i
            \end{array}\right.$

		    \State \textit{reward} = $Env.\Call{step}{v_t}$
		    \State Compute a $target\_value$ using \textit{reward} on GPU$^i$

            \State $/*$ Updates graph $*/$
		    \State $S^i$ += $v_{t}$, $C^{i}$ -= $v_{t}$, revise $A^i$ based on $v_t$; // all occur on GPU$^i$

		    \State $/*$ Add a new tuple to replay buffer $*/$
		    \State Replay\_Buffer += (index of $g$, $S^i$, $v_t$, target\_value)
		    
    		\State $/*$ Randomly sample a batch of $B$ tuples using same seed $*/$
		    \State tuples\_batch = \Call{sample}{Replay\_Buffer, size=B}
		    
	        \State $/*$ Lines 21-24: process the tuple\_batch $*/$
	        \State $/*$ Create an adjacency matrix for each subgraph in \par 
	        \hskip\algorithmicindent \quad tuples\_batch, then stack them up in 3D $*/$ 
	        \State $batched\_A^i$ = \Call{Tuples2Graphs}{list of graph index \par
	        \hskip\algorithmicindent \quad \ \ and $S^i$ in tuples\_batch}

	        \State $batched\_S^i$ = \Call{stack}{list of $S^i$ in tuples\_batch}
	       
	        \State $batched\_C^i$ = \Call{stack}{list of $v_t$ in tuples\_batch}
	        
   	        \State $batched\_target\_values^i$ = \Call{stack}{list of \par
	        \hskip\algorithmicindent \quad \ \ target\_values in tuples\_batch}
	        
		    \State $/*$ Apply multiple iterations to train $EM$ and $Q$ $*/$

		    \State \Call{Train}{EM, Q, $batched\_A^i$, $batched\_S^i$,
		    \par \hskip\algorithmicindent \quad \ \ 
		    $batched\_target\_values^i$} on GPU$^i$
		    
		    \If{All edges are covered in graph ${g}$} break
		    \EndIf
			
			\EndFor
			\EndFor 
			
		\end{algorithmic}
	}
\end{algorithm}

\begin{figure} [t] 
    \centering
	\includegraphics[width=0.5\textwidth]{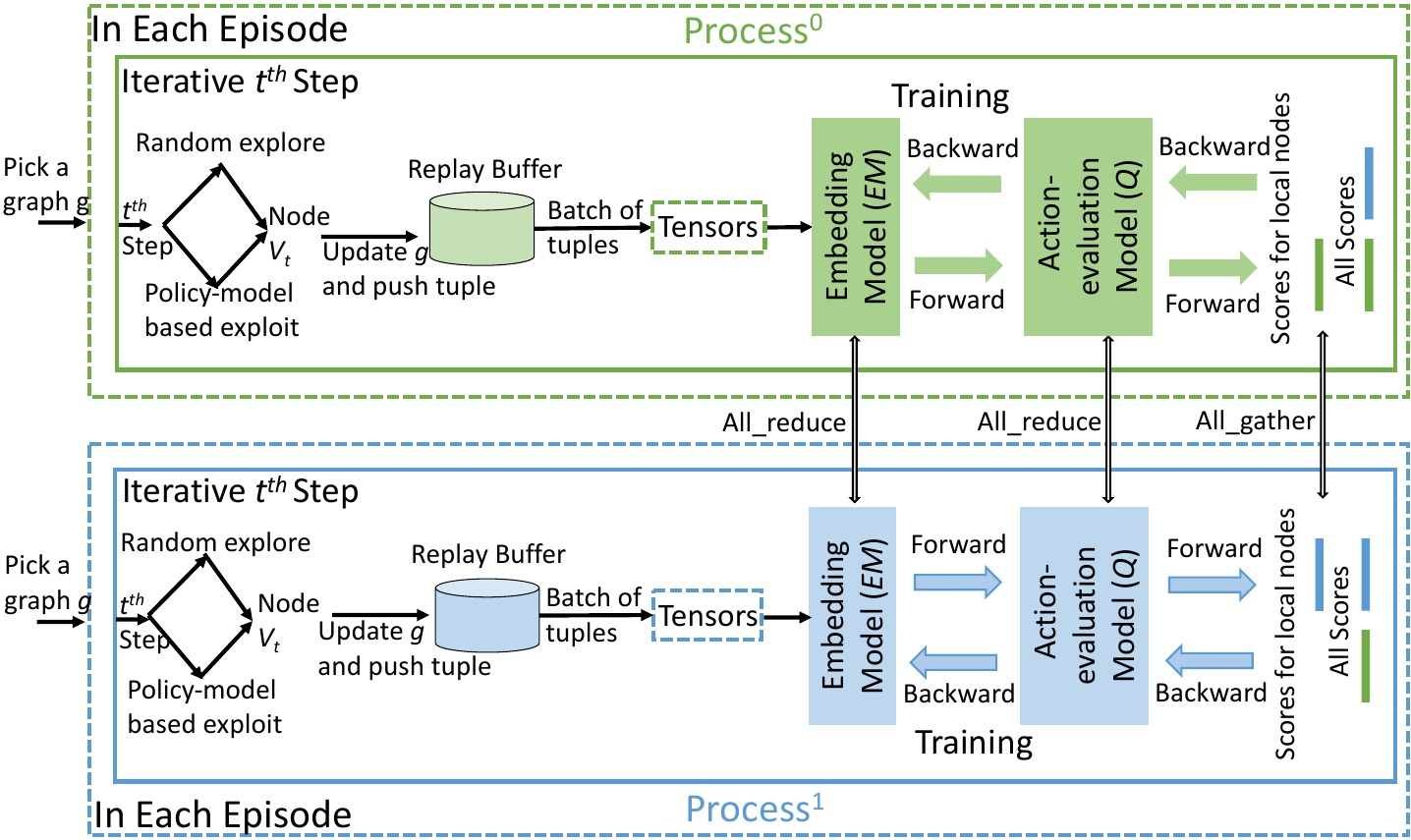}
	\par \centering
	\caption{\small{
	A single step of the parallel training algorithm (Alg. \ref{alg:parallel_rl_training}) in OpenGraphGym-MG. At the beginning of each episode, the agent on each process picks a training graph from the training graph dataset. Then, each agent selects a node $V_t$ randomly or through policy model ($EM$ and $Q$). Next, the agent updates the corresponding local adjacency matrix $A^i$, the local partial solutions $S^i$, and the local set of candidate nodes $C^i$. The agent then form a new tuple and pushed it to the replay buffer. Next, the agent samples a mini-batch of tuples and transforms them into 3D tensors of $A^i$, $S^i$, $C^i$, and the target values. Finally, the agent applies multiple gradient descent steps to train the policy model ($EM$ and $Q$). 
	}
	}
	\label{fig:par_train}
\end{figure}

Proc$^i$ starts a new episode by randomly selecting a graph $g$ from the training graph dataset. We use the same seed among all processes so that the graph selected by all processes is the same.
Each new episode corresponds to a different graph problem instance.
Given the selected graph $g$,
each process creates an environment $Env$ for the graph.
To solve the graph instance $g$ (in the episode), Proc$^i$ begins by either selecting a node randomly (i.e., explore), or using the current policy model (i.e., exploit) to decide the ``best'' node.
After the best node $v_t$ is decided, the graph $g$'s state ($S^i$, $C^i$, and $A^i$) will be updated to reflect that $v_t$ changes from a candidate node to a solution node.
Next, a new experience tuple is appended to the Replay Buffer,
which is of (index of graph $g$, $S^i$, $v_t$, target\_value).
Starting from Line 18, all processes will carry out the distributed training step, in which
Proc$^i$ samples a mini-batch of tuples from its Replay Buffer, and launches the training function.
A single step of executing Alg. \ref{alg:parallel_rl_training} on two processes is illustrated in Fig. \ref{fig:par_train}. 

{\it Optimization of Replay Buffer to Reduce Memory Cost:}
Since a replay buffer often contains tens of thousands of tuples,
it is too expensive to store a graph's current state (such as its adjacency matrix) in each experience tuple.
Hence, we develop a method that only stores a graph's index and its partial solution to minimize the memory space.
However, we still need to restore the graph's state and adjacency matrix before training starts.
To do that, each process uses the graph's current partial solution and the original graph's adjacency matrix 
to generate a subgraph dynamically. 
This functionality is realized by the 
\Call{Tuples2Graphs}{} function (Line 21 of Alg. \ref{alg:parallel_rl_training}), which converts each tuple's information to a concrete adjacency matrix.
A number of $B$ tuples will result in a 3D tensor
(i.e., a stack of adjacency matrices).
Lines 22-24 will stack the lists of $S^i$, $v_t$, and target\_value from $B$ tuples into three more 3D tensors.
Finally, each GPU trains the EM and Q models by using the prepared 3D tensors of $batched\_A^i$, $batched\_S^i$, $batched\_C^i$, and $batched\_target\_values$. 


{\it Implementation details of the \Call{Train}{} function:}
The forward propagation of the \Call{Train}{} function essentially executes Alg.~\ref{alg:3d_parallel_em_model} then Alg.~\ref{alg:3d_parallel_q_model}. 
We use {\it PyTorch Distributed} \cite{pytorch} to implement the forward propagation, and use the Adam optimizer \cite{adam_optimizer} to train the policy model. 
The backward propagation 
is implemented by using the Adam algorithm provided by the PyTorch {\it optim} package~\cite{pytorch-optim} 
via calling the functions of \textit{loss.backward()} and \textit{optimizer.step()}.

%% file: sections/4_5_optimizations.tex
\subsection{Additional Optimization Techniques}
\label{subsec:mg_optimizations}

In the OpenGraphGym-MG framework, we design and develop two more performance optimization techniques to further accelerate the RL inference process and training process.

\subsubsection{Multiple-Node Selection per Policy Evaluation}~

The original RL inference algorithm (Alg. \ref{alg:parallel_rl_testing}) 
may take up to $|V|$ steps (i.e., $|V|$ times of the policy model evaluations) to solve a large-scale graph optimization problem. 
At each step, the node with the highest score is selected to become a part of the solution.
To speed up the essentially ``sequential'' process, we design an adaptive node-selection strategy that, after $P$ GPUs collectively compute scores for all the candidate nodes, they select $d$ nodes with the top $d$ scores as a subset of the partial solution (Note: the original RL inference algorithm is the special case when $d = 1$). 
 This design is based on the observation that, when the size of a graph is big enough, the set of $d$ nodes selected sequentially (through $d$ steps) is similar to the set of $d$ nodes with the top $d$ scores (at the beginning of the $d$ steps). 
This way, the RL agent is able to select $d$ nodes based on one evaluation of the policy model, potentially delivering a speedup of $d$ times.


However, selecting a large $d$
may also result in a degraded quality of solution.
Therefore, we use an adaptive scheme to gradually decrease the number of selected nodes $d$ per step:
When the current set of candidate nodes $|C|$ is larger than $\frac{N}{2}$, $d$ is equal to 8. 
$|C|$ will become smaller as more nodes are added to the partial solution.
As $|C| \in(\frac{N}{4}, \frac{N}{2}]$, $d$ is lowered to 4. 
As $|C| \in(\frac{N}{8}, \frac{N}{4}]$, $d$ is lowered to 2. 
When $|C|$ becomes less than $\frac{N}{8}$, $d$ returns to 1.
The basic idea is that when there are many candidate nodes to select, we try to be aggressive, and gradually become more conservative when there are fewer nodes left. We show the performance of this optimization technique in Section \ref{subsec:opt_exp}. 

\subsubsection{Optimizing the Number of Gradient-Descent Iterations}~

In the original RL training algorithm, an agent always samples a mini-batch of $B$ tuples
from the Replay Buffer, 
and then executes one iteration of the gradient-descent training (i.e., one forward propagation followed by one backward propagation).
Instead of performing only one iteration of the gradient descent, we tune the number of gradient-descent iterations for the mini-batch to improve the convergence rate.
Based on our experiments, we find that the agent with multiple iterations of gradient descent can 
converge to an optimized policy model faster.
In other words, the RL agent is able to learn an optimized policy model using fewer training steps (i.e., having a faster convergence rate).
Nevertheless, a too large number of gradient-descent iterations can also cause the RL training process to become more unstable.
In Section \ref{subsec:opt_exp}, we show the experimental results using different numbers of gradient-descent iterations.

%% file: sections/5_analysis.tex
\section{Analysis of OpenGraphGym-MG}
\label{sec:analysis}


\subsection{Parallel Efficiency Analysis}
\label{subsec:time_complexity}


{\bf (1) Efficiency analysis of the $EM$ model evaluation
on $P$ GPUs:}
The Alg.~\ref{alg:3d_parallel_em_model} of graph embedding evaluation can be configured with $L$ recurrent embedding layers and an embedding dimension $K$. 
We assume each graph has $N$ nodes with an edge probability of $\rho$. 

The time complexity of the parallel graph embedding computation using $P$ GPUs can be expressed as follows:   

\small
\begin{equation}
\label{equa:embed_par}
\begin{split}
T^{}_{embed}(B,N,\rho;K,L;P)   
&=  \frac{N^2}{P}(BK(\rho + L) + \frac{BK(2+K+4L)}{N}) \\
     &+ \alpha L \log_2 P  + \beta LBKN,
\end{split}
\end{equation}
\normalsize

where $\alpha$ is the network latency and $\beta$ is the reciprocal of the network bandwidth. Note that there are a number $L$ of MPI\_All\_reduce communications, and each message is of size $B\times K \times N$.

Time complexity of the sequential algorithm is:
\small
\begin{equation}
\label{equa:embed_seq}
\begin{split}
T^{}_{embed\_seq}(B,N,\rho;K,L) &= N^2(BK(\rho + L)+ \frac{BK(2+K+4L)}{N})
\end{split}
\end{equation}
\normalsize

Hence, the Parallel Efficiency of
$E_{embed}(P)$ = $(\frac{T_{embed}(P)}{T_{embed\_seq}/P})^{-1}$ 
$\approx$ 
$(1+\frac{\beta P}{N(1+\frac{\rho}{L})})^{-1}$, which is close to 1.0 when 
$P$ is much less than $N$.

{\bf (2) Efficiency analysis of the action-evaluation model on $P$ GPUs:}
Alg. \ref{alg:3d_parallel_q_model} computes the scores for all the candidate nodes based on the embeddings produced from the $EM$ model.
The time complexity of the parallel action-evaluation model computation using $P$ GPUs is expressed as follows:

\small
\begin{equation}
\label{equa:eva_par}
T^{}_{action}(B,N,\rho;K,L;P) = \frac{BKN}{P}(6 + K + \frac{KP}{N})  + \alpha \log_2 P  + \beta BK
\end{equation}
\normalsize
Note that the action-evaluation algorithm has one MPI\_All\_reduce communication, and each message is of size $B\times K$.

Time complexity of the sequential algorithm is:
\small
\begin{equation}
\label{equa:eva_seq}
T^{}_{action\_seq}(B,N,\rho;K,L) = BKN(6 + K + \frac{K}{N})
\end{equation}
\normalsize

Hence, the parallel efficiency of the action evaluation model is: 
\small 
\begin{equation}
\label{equa:eva_eff}
E_{action}(P) 
 = (\frac{T_{action}(P)}{T_{action\_seq}/P})^{-1} 
 \approx (1 + \frac{P}{cN+1} + \frac{\beta}{N(K+6)})^{-1},
\end{equation}
\normalsize

where $c = \frac{K+6}{K}$. Since $N \gg P$, Parallel Efficiency $E^{}_{action}(P)$ is almost equal to $1.0$. Here, we skip the computation cost analysis for the backward propagation ~\cite{time_complexity,  li2014highly} whose complexity is similar to that of the forward propagation.

{\bf (3) Analysis of the computation cost on the host:}
In RL training, we use GPUs to compute the evaluation of the policy model and training of the policy model, as shown in Alg.~\ref{alg:parallel_rl_training}.
In addition to using GPUs, we also use multiple CPUs on the host to simulate the graph problem-solving environment, and generate data for RL training.
The operations computed on the host include: getting a reward signal, sampling experience tuples, updating local graph data structures, and generating training data from tuples.
Among them, only the two
operations of updating local graph data structures (Line 14 of Alg.~\ref{alg:parallel_rl_training}) and generating an adjacency matrix tensor (Line 21) have a $>O(1)$ time complexity, which are $\frac{2\rho N}{P}$ and $\frac{2\rho N^2 B}{P}$, respectively.
However, it is easy to see
that the host-based computation has a parallel efficiency 
that is close to 1 due to the algorithm design using $P$ processes concurrently.

Besides, the RL inference Alg. \ref{alg:parallel_rl_testing} 
has an MPI\_All\_gather communication at each step to collect all scores from $P$ GPUs, in which each GPU sends a vector of $N/P$ floating numbers. 
Regarding the backward propagation for training the $EM$ and $Q$ models,
its communication cost is a global reduction of the gradients of the model parameters: $\theta_1-\theta_7$ (i.e., $4K^2 + 4K$ floating point numbers).

\subsection{Memory Cost Analysis}
\label{subsec:memory_usage}
In RL training, a mini-batch of $B$ graphs corresponds to an adjacency matrix tensor $A\in \mathbb{R}{}^{B \times N \times N}$, a partial solution tensor $S\in \mathbb{R}{}^{B \times N \times 1}$, and a tensor of candidate nodes $C\in \mathbb{R}{}^{B \times N \times 1}$, where $N$ is the number of nodes in each graph. Since we distribute each data structure across $P$ GPUs, every GPU on average stores $\frac{1}{P}$-th of the $A$, $S$, and $C$ tensors, respectively.

In our framework, each adjacency matrix is stored in the sparse COO (Coordinate) format,
using {\tt torch.sparse.FloatTensor}. 
The sparse COO format only stores the non-zero elements. 
The amount of memory required for storing one adjacency matrix is $\frac{20N^2\rho}{P}$ bytes on each GPU, given edge probability of $\rho$. Note that $N^2\rho$ is equal to the number of edges.
For a mini-batch of $B$ graphs, the adjacency matrix tensor takes $\frac{20N^2\rho B}{P}$ bytes of memory on each GPU.

In addition, the number of bytes needed to store
partial solutions and sets of candidate nodes are
each $\frac{4NB}{P}$ bytes.
Also, suppose the Replay Buffer consists of $R$ experience tuples,
then it takes 
$8R(\frac{N}{P}+1)$
bytes to store $R$ experience tuples on each GPU.

%% file: sections/6_experiment.tex
\section{Evaluation}
\label{sec:exp}

To demonstrate the effectiveness and efficiency of OpenGraphGym-MG, we conduct
three types of experiments: 
1) to evaluate the learning speed and solution quality, 
2) to investigate the effect of the additional optimization methods, and 
3) to demonstrate the scalability of the parallel RL algorithms on large-scale graphs using multiple GPUs.

\subsection{Experimental Setup}
\label{subsec:exp_setup}

{\bf Software:} 
With respect to software, we use the PyTorch library 1.3.1~\cite{pytorch}, graph library NetworkX 2.3~\cite{networkx}, and linear programming library IBM-CPLEX 12.10~\cite{cplex}. 
For data communication between different GPU-equipped processes, 
we use the collective communication library from {\it torch.distributed}, whose backend is the Nvidia library NCCL~\cite{nccl}. 
We use NetworkX to generate graphs. 
To evaluate the quality of solution computed by the RL framework,
we use the IBM-CPLEX Solver to obtain a reference optimal solution, for which
the IBM-CPLEX solution-to-time cutoff is set to 0.5 hours~\cite{ilp}. 

{\bf Hardware:} 
All experiments are conducted on the Summit supercomputer at the Oak Ridge National Laboratory (ORNL). 
Each Summit node (an IBM Power System AC922) contains two Power9 CPUs, and six 16GB Nvidia V100 GPUs. 
There are 44 CPU cores and 512 GB DDR4 memory in each node.

{\bf Datasets:} We use Erd{\H{o}}s-R{\'e}nyi (ER) \cite{er_graph} graphs and Barab{\'a}si-Albert (BA) \cite{ba_graph} graphs as well as real-world social network graphs~\cite{real_world_graph1} to do experiments. 
The generation of ER graphs is controlled by a model $ER(n, \rho)$, in which $n$ is the number of the nodes, and each pair of nodes has a possibility of $\rho$ to be connected with an edge. 
We set $\rho$ to 0.15 to generate ER graphs. 
For the BA graph model $ER(n, d)$, it generates a graph by adding each node to the existing graph. For each new node, there will be $d$ edges connected from the new node to the existing nodes. We set $d=4$ to generate BA graphs. 
As to the real-world graphs, we use three universities' Facebook friendship networks \cite{facebook_network}. We obtain the data from NetworkRepository \cite{real_world_graph1}. The graphs'
information is summarized in Table \ref{tbl:real_graph_info}.

\begin{table}[h]
    \renewcommand{\arraystretch}{1.2}
	\centering
	\caption{\small Information of the real world graphs~\cite{real_world_graph1}.}
	\label{tbl:real_graph_info}
	\small{
		\begin{tabular}{|  c  |  c  |  c  |  c  |}
        \hline
        Dataset name & $|V|$ & $|E|$  & Edge probability $\rho$ \\
        \hline
        \hline
        Vanderbilt & 8.1K & 427.8K & 0.0131 \\
        \hline
        Georgetown & 9.4K & 425.6K & 0.0096 \\
        \hline
        Mississippi & 10.5K & 610.9K & 0.0110 \\
        \hline
        \end{tabular}
	}
\end{table}

{\bf Hyper-parameter Settings:} In our experiments, the RL exploration rate ($\epsilon$) is set as a decayed rate that decreases from 0.9 to 0.1. The learning rate ($\eta$) used to train the policy model is \num{1.0e-5}. The size of the Replay Buffer is set to 50,000. The discount factor ($\gamma$) for Bellman Equation is set to 0.9. The number of embedding layers ($L$) and the dimension of graph embedding ($K$) are set to 2 and 32, respectively.

\subsection{Evaluation of Learning Speed}
\label{subsec:converg_exp}


\begin{figure} [th] 
	\centering
	\begin{subfigure}[h]{0.25\textwidth}
	\centering
		\includegraphics[width=\linewidth, height=.725\textwidth]{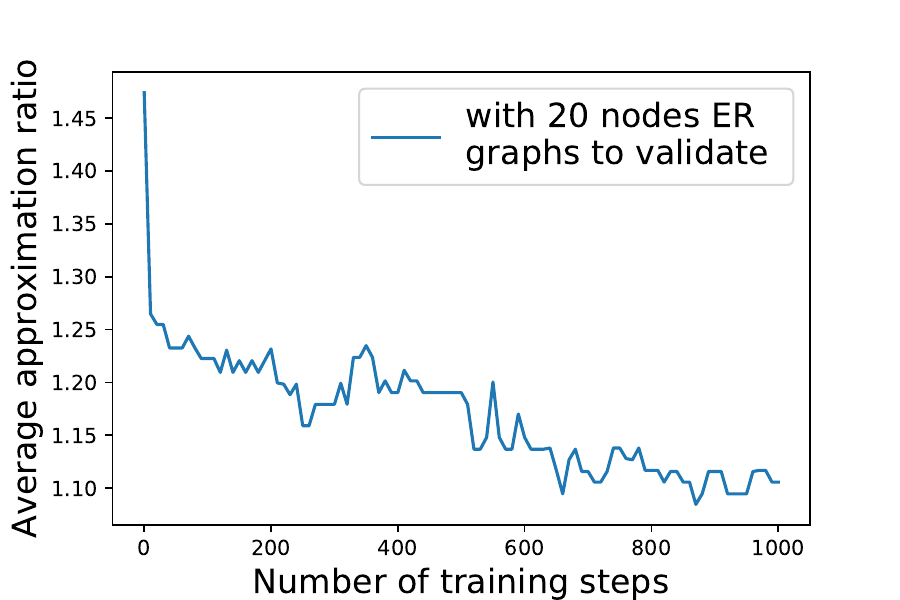}
		\par \centering {\footnotesize (1a)}
	\end{subfigure}
	~
	\begin{subfigure}[h]{0.25\textwidth}
	\centering
    	\includegraphics[width=\linewidth, height=.725\textwidth]{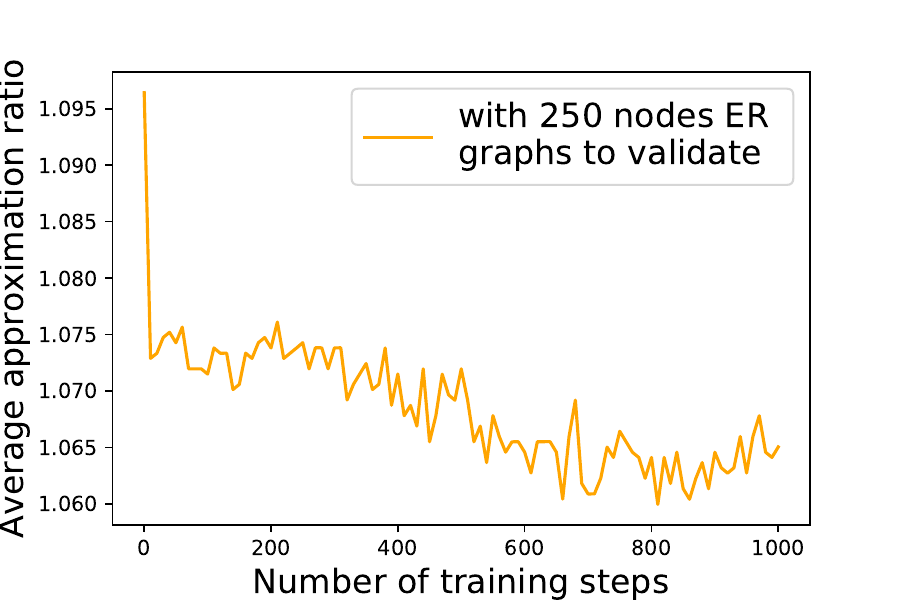}
    	\par \centering {\footnotesize (1b)}
    	\end{subfigure}
	\hfill
	
	\centering
	\begin{subfigure}[h]{0.25\textwidth}
	\centering
		\includegraphics[width=\linewidth, height=.725\textwidth]{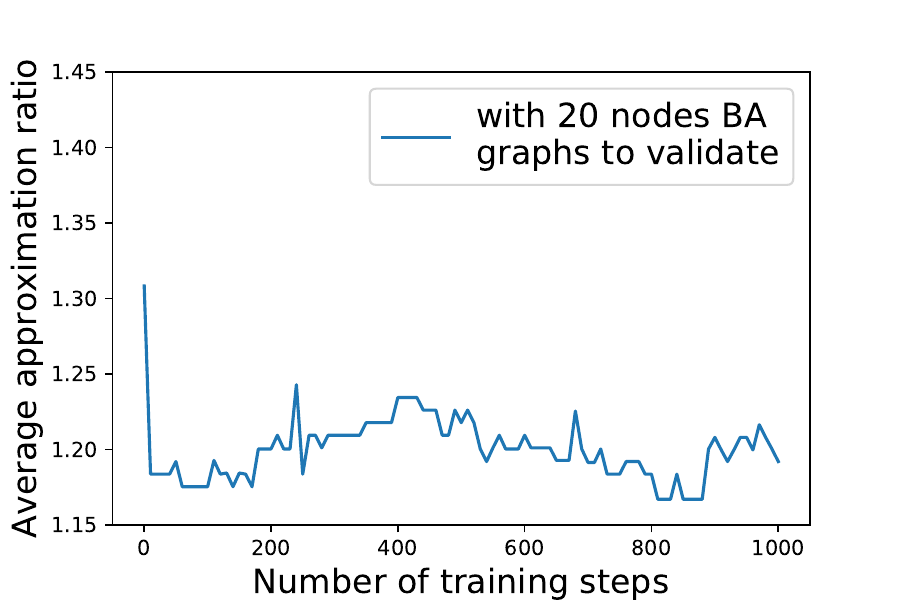}
		\par \centering {\footnotesize (2a)}
	\end{subfigure}
	~
	\begin{subfigure}[h]{0.25\textwidth}
	\centering
    	\includegraphics[width=\linewidth, height=.725\textwidth]{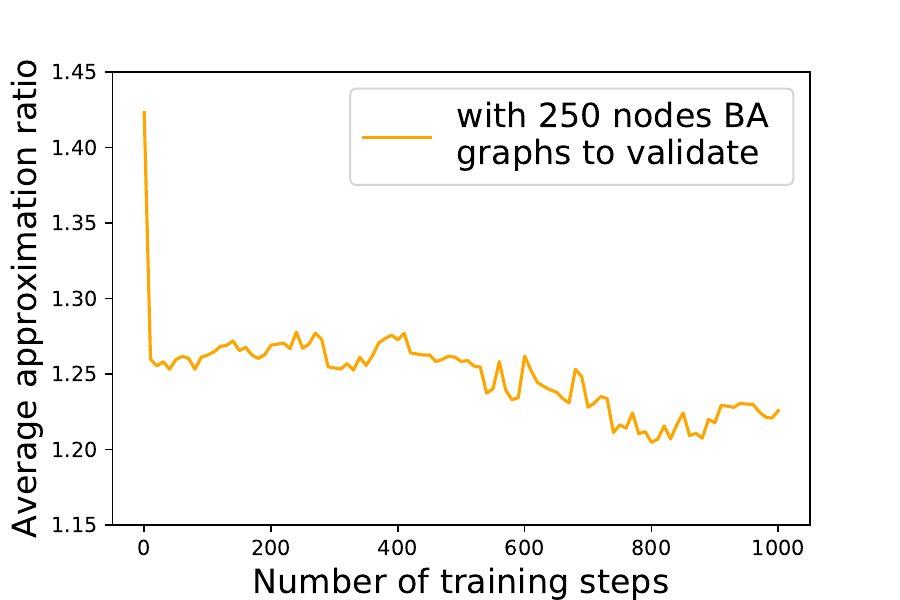}
    	\par \centering {\footnotesize (2b)}
    	\end{subfigure}
	\hfill
	
	\caption{\small
	Evaluation of the RL learning speed with ER graphs (1a \& 1b) and BA graphs (2a \& 2b).
	}
	\label{fig:convergence}
\end{figure}

We perform experiments to measure how well our graph RL framework will operate on the MVC graph problems in terms of learning speed (i.e., the performance over training by periodically testing the solution quality of a fixed set of test graphs).
Fig. \ref{fig:convergence} shows the learning curves of our RL agent trained on both ER and BA graphs on one GPU.
We use graphs of size $|V|=20$ to train, and use graphs of two different sizes ($|V|=20$ and $|V|=250$) to test.
To measure the learning curve, 
we collect the average {\it approximation ratio} of the 10 test graphs' solutions every 10 RL training steps. 

In Fig. \ref{fig:convergence} 1a and 1b, the RL agent
is trained with 20-node ER graphs, and tested with 10 test graphs.
In subfigure 1a, the test graphs we used have 20 nodes in each graph .
We can see that the RL agent
improves the average approximation ratio from 1.5 to 1.1 after 1000 training steps. 
In subfigure 1b, the test graphs have 250 nodes per graph
although the training graph only has 20 nodes.
Subfigure 1b shows that the agent can still improve the average approximation ratios quickly after 1000 steps. 

With respect to the BA type of graphs, Fig. \ref{fig:convergence} 2a and 2b display a similar fast learning curve, where the average approximation ratio reduces from 1.32 to 1.17 on test graphs with 20 nodes, and 1.43 to 1.2 on test graphs with 250 nodes. 
Results in Fig. \ref{fig:convergence} empirically show that the RL agent can generalize to test graphs with a larger number of nodes. 


\subsection{Effect of Two Optimization Techniques}
\label{subsec:opt_exp}

{\it 1) Effect of the Multiple-Node Selection Technique:}~

Based on the multiple-node selection strategy described in
Section \ref{subsec:mg_optimizations},
our optimized RL inference algorithm can select multiple nodes to  add to the partial solution at every step.
Here, we do experiments to compare the original inference algorithm with the multiple-node selection method.
In the experiments, we measure the total time for
a pretrained agent to search for optimal MVC solutions, given
three unseen test graphs with 750, 1500, and 3000 nodes, respectively.
Note that our multiple-node selection method employs the adaptive scheme specified in Section \ref{subsec:mg_optimizations}. 

\begin{figure} [t] 
    \centering
	\includegraphics[width=0.36\textwidth] 
	{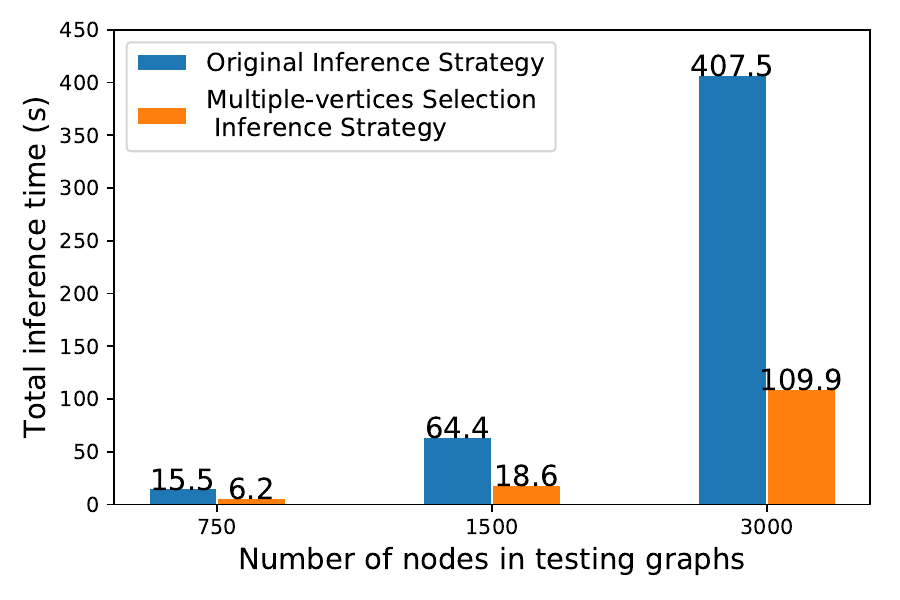}
	\par \centering
	\caption{\small
	Comparison between the original RL inference algorithm (d=1) and the algorithm that uses the adaptive multiple-node selection technique.}
	\label{fig:multiple_vertices}
\end{figure}

Fig. \ref{fig:multiple_vertices} shows the comparison
between the original inference algorithm (i.e., selecting one node per step) and the optimized multiple-nodes selection method.
Given a graph with 750 nodes, the original inference algorithm takes 15.5 seconds while the optimized algorithm takes 6.2 seconds.
If we compare the two solutions found by the two algorithms: $MVC_{orig}$ and $MVC_{new}$, their ratio $\frac{|MVC_{new}|}{|MVC_{orig}|}=1.008$, which is significantly close.

When the test graph size increases to 1,500 nodes,
the original algorithm takes 64.4 seconds and the optimized algorithm takes 18.6 seconds, which is 3.5 times faster. 
Their MVC solutions have a ratio of $\frac{|MVC_{new}|}{|MVC_{orig}|}=1.002$.
For a test graph with 3,000 nodes,
the optimized inference algorithm is 3.7 times faster than the original algorithm (i.e., 109.9 versus 407.5 seconds). Their MVC solution ratio $\frac{|MVC_{new}|}{|MVC_{orig}|}=1.004$.
Based on the experiments, if the graph size is large, selecting a relatively small number of nodes per inference step will not significantly degrade the quality of the solution. 


\begin{figure} [b] 
    \centering
	\includegraphics[width=0.45\textwidth] 
	{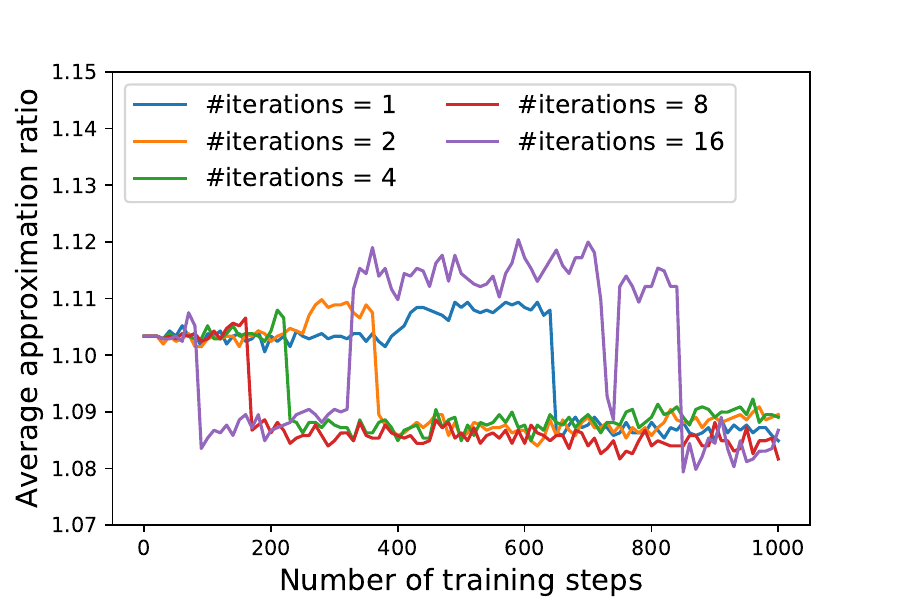}
	\par \centering
	\caption{\small Impact of the gradient-descent iterations on the solution. 
	}
	\label{fig:local_iter}
\end{figure}

{\it 2) Effect of the Number of Gradient-Descent Iterations:}~ 

The other optimization technique introduced
in Section \ref{subsec:mg_optimizations} is to tune
the number of gradient-descent iterations 
$\tau$ to speed up the RL training process.
To evaluate the effect of the optimization, we test multiple numbers of gradient-descent iterations from 1 to 16.
In the experimental results shown in Fig. \ref{fig:local_iter},
we use a set of graphs with 250 nodes to train the agent, and periodically test the solution quality on 10 unseen 250-node graphs every 10 training steps. 

From Fig. \ref{fig:local_iter}, we can see
that the original RL algorithm (i.e., $\tau=1$) can converge to a solution quality with an average approximation ratio of $1.08$ after around 650 steps. 
As $\tau$ increases to 2, the RL algorithm can converge to a similar solution of quality after around 400 steps, which shows a faster learning speed.
Similarly, increasing $\tau$ to 4 and 8 continues to improve the learning speed to obtain the same average approximation ratio, 
using around 230 and 200 steps, respectively.
However, $\tau$ cannot be too big.
When $\tau$ is equal to 16, 
we start to see significant oscillation in the learning curve. 
Although the results empirically show it can be beneficial to increase the number of gradient-descent iterations, a theoretical analysis is still needed to study the effect of tuning this parameter.

\subsection{Scalability of OpenGraphGym-MG}
\label{subsec:sca_exp}

In addition to the theoretical performance analysis of the parallel RL inference and training algorithms (shown in Sections \ref{subsec:mg_inference} and \ref{subsec:mg_training}),
we also perform experiments to measure the actual scalability performance of OpenGraphGym-MG on multiple GPUs.
We perform the experiments with different ER graphs and real-world graphs using between 1 and 6 Nvidia Volta GPUs. 

\begin{figure} [b] 
    \centering
	\includegraphics[width=0.37\textwidth] 
	{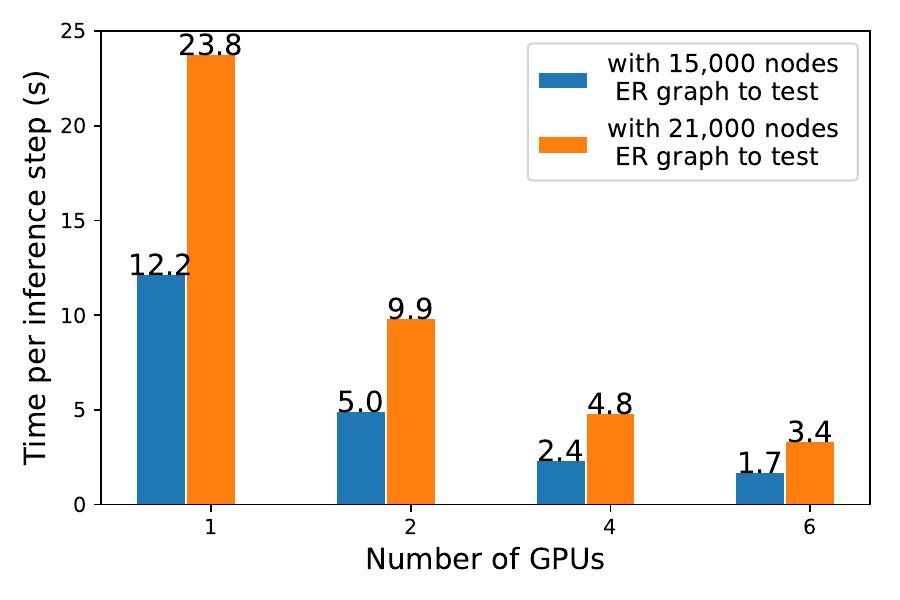}
	\par \centering
	\caption{\small Execution time of a single parallel RL inference step over large ER graphs.
	}
	\label{fig:test_sca}
\end{figure} 

\begin{figure} [b] 
    \centering
	\includegraphics[width=0.37\textwidth] 
	{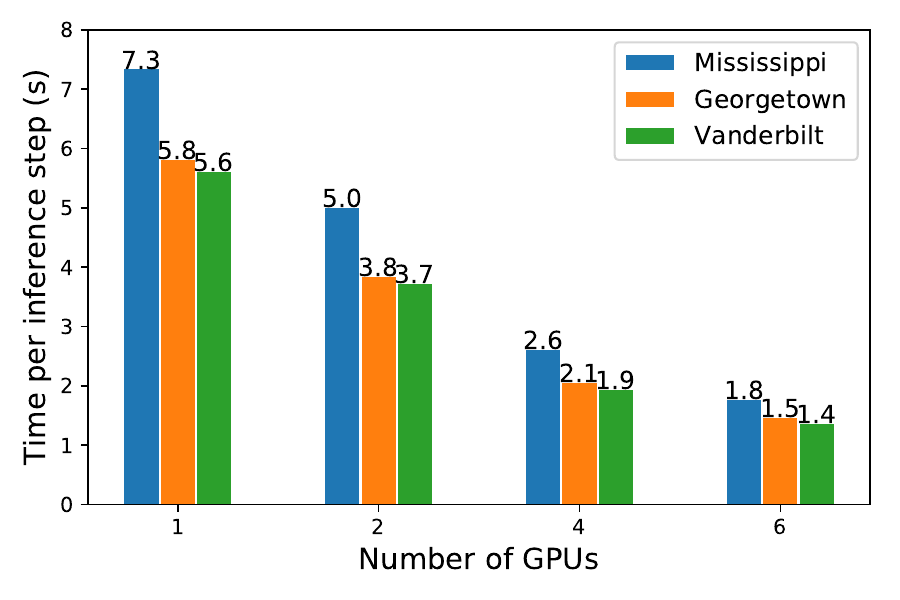}
	\par \centering
	\caption{\small Execution time of a single parallel RL inference step over large real-world graphs. 
	}
	\label{fig:test_sca_real}
\end{figure} 

Fig. \ref{fig:test_sca} shows the performance of the parallel RL inference using OpenGraphGym-MG. 
Here we report the execution time per inference step.
For RL inference, the time-per-step is defined as the time to invoke the agent's policy model, and select a node to add to the partial solution. 
Based on Fig. \ref{fig:test_sca},
when the graph size is 15,000, the average time per step is reduced 
from 12.2s to 1.7s by using 6 GPUs.
When the graph size is 21,000, the average time per step 
can be reduced from 23.8s to 3.4s (around 7 times faster). 
Our later performance profiling reveals that
the super-linear speedup is caused by the specific single-GPU experiment taking a longer time to update the adjacency matrices due to the big graph size.

As to applying RL inference to real-world graphs, OpenGraphGym-MG can obtain a speedup of 4.1 times on 6 GPUs, as shown in Fig.~\ref{fig:test_sca_real}.
We notice the speedup using the real-world graph dataset is less than that using the large ER graph dataset (in Fig. \ref{fig:test_sca}).
This is because the total amount of computation is relevant to the number of edges, and the real-world graphs have much fewer edges than the ER graphs with tens of millions of edges.



\begin{figure} [thb] 
    \centering
	\includegraphics[width=0.37\textwidth] 
	{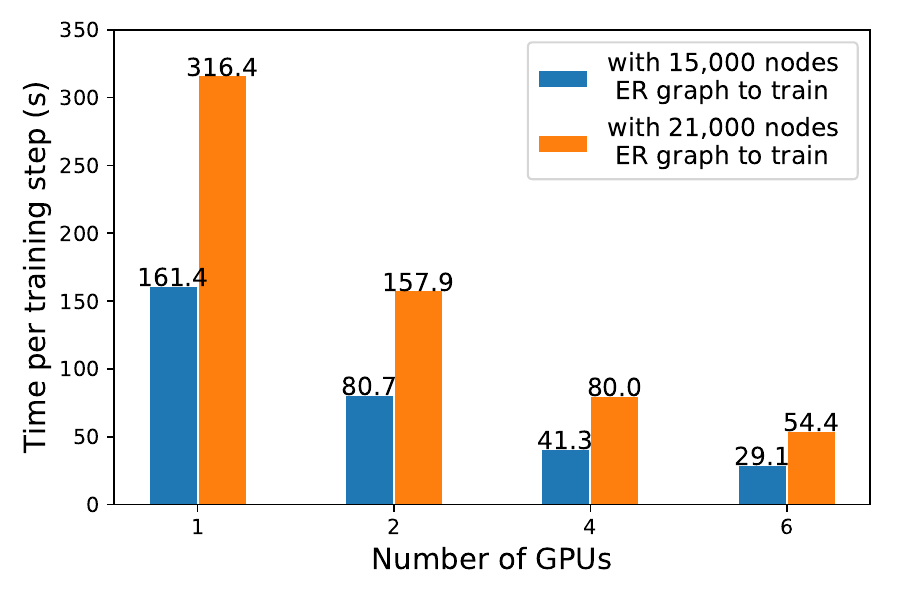}
	\par \centering
	\caption{\small Execution time of a single parallel RL training step over large ER graphs. 
	}
	\label{fig:train_sca}
\end{figure}

In Fig. \ref{fig:train_sca}, we show the performance of the parallel RL training using OpenGraphGym-MG.
We measure the time per training step in the experiment. 
The experimental results in Fig.~\ref{fig:train_sca} show that as the number of GPUs increases, 
for graphs with 15,000 nodes, the framework can reduce the time per training step from 161.4s to 29.1s (i.e., 5.5 times faster) on 6 GPUs.
As for larger graphs with 21,000 nodes (having around 33 million edges),
the time per training step can be reduced from 316.4s to 54.4s (i.e., 5.8 times faster) by using 6 GPUs.
Overall, both the parallel inference results (Fig. \ref{fig:test_sca} and Fig. \ref{fig:test_sca_real})
and parallel training results (Fig. \ref{fig:train_sca})
show good scalability on the Summit supercomputer. 

%% file: sections/7_conclusion.tex
\section{Conclusion}
\label{sec:conclusion}

In this paper, we presented a high performance OpenGraphGym-MG framework to solve large-scale graph optimization problems on multiple GPUs. 
Using the Minimum Vertex Cover (MVC) as an example, we introduced the parallel RL inference and training algorithms and implementations in OpenGraphGym-MG.
We also developed several optimization techniques to further optimize the RL inference and training performance.
The theoretical parallel efficiency analysis and memory cost analysis proved the parallel RL inference and training algorithms are efficient and scalable on a number of GPUs.
The experimental results showed that our optimized graph RL framework reduced the inference and training time significantly on multiple GPUs when it applied to large graphs of more than 30 million edges.    
Our future work will extend OpenGraphGym-MG to solve extreme-scale graph problems on a distributed-memory cluster with many MultiGPU nodes.